\documentclass[journal=jctcce, manuscript=article]{achemso}
\usepackage{amsmath}
\usepackage{xspace}
\usepackage{multirow}
\usepackage{latexsym}
\usepackage{braket}
\usepackage{cleveref}
\usepackage{threeparttable}
\usepackage{enumerate}
\usepackage{mathrsfs}
\usepackage{enumitem}
\usepackage{color}
\usepackage[version=3]{mhchem}
\usepackage{caption,setspace}
\usepackage{graphicx}
\usepackage{subfig}
\usepackage{threeparttable}
\usepackage{bbold}

\allowdisplaybreaks
\usepackage{array}
\newcolumntype{L}[1]{>{\raggedright\let\newline\\\arraybackslash\hspace{0pt}}m{#1}}
\newcolumntype{C}[1]{>{\centering\let\newline\\\arraybackslash\hspace{0pt}}m{#1}}
\newcolumntype{R}[1]{>{\raggedleft\let\newline\\\arraybackslash\hspace{0pt}}m{#1}}

\crefname{figure}{Figure}{Figures}
\crefname{table}{Table}{Tables}
\crefname{equation}{Eq.}{Eqs.}
\crefname{section}{Section}{Sections}

\usepackage[defaultcolor=red]{changes}
\definechangesauthor[name={Koushik}, color=red]{KC}
\title{Analytic evaluation of non-adiabatic couplings within 
the complex absorbing potential equation-of-motion coupled-cluster 
method}

\author{Koushik~Chatterjee}
\email{koushikchatterjee7@gmail.com}
\affiliation{Department of Chemistry, KU Leuven, Celestijnenlaan 200F, 
B-3001 Leuven, Belgium}

\author{Zsuzsanna Koczor-Benda}
\affiliation{Department of Chemistry, University of Warwick, Coventry, 
CV4 7AL, UK}

\author{Xintian Feng}
\affiliation{Q-Chem, Inc., 6601 Owens Drive, Suite 240, Pleasanton, 
CA 94588, USA}

\author{Anna I. Krylov}
\affiliation{Department of Chemistry, University of Southern California, 
Los Angeles, CA 90089, USA}

\author{Thomas-C.~Jagau}
\affiliation{Department of Chemistry, KU Leuven, Celestijnenlaan 200F, 
B-3001 Leuven, Belgium}

\begin{tocentry}
{\includegraphics[width=\textwidth]{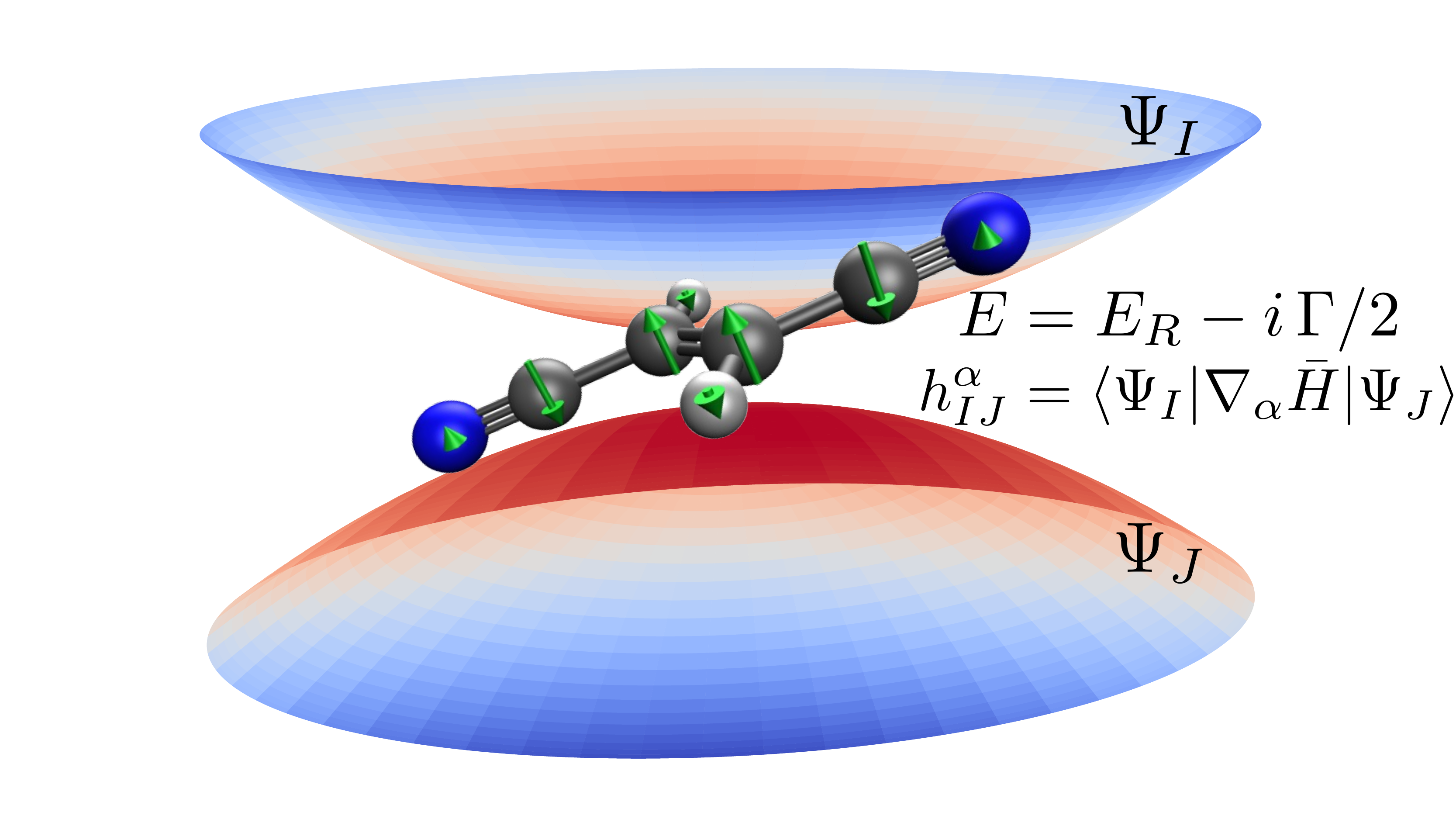}}
\end{tocentry}

\begin{document}
%\preprint{APS/123-QED}

\begin{abstract}
We present the theory for the evaluation of non-adiabatic couplings 
(NACs) involving resonance states within the complex absorbing potential 
equation-of-motion coupled-cluster (CAP-EOM-CC) framework implemented 
within the singles and doubles approximation. Resonance states are 
embedded in the continuum and undergo rapid decay through autodetachment. 
In addition, nuclear motions can facilitate transitions between different 
resonances and between resonances and bound states. These non-adiabatic 
transitions affect the chemical fate of resonances and have distinct 
spectroscopic signatures. The NAC vector is a central quantity needed 
to model such effects. 

In the CAP-EOM-CC framework, resonance states are treated on the same 
footing as bound states. Using the example of fumaronitrile, which supports 
a bound radical anion and several anionic resonances, we analyze the 
non-adiabatic coupling between bound states and pseudocontinuum states, 
between bound states and resonances and between two resonances. We find 
that the NAC between a bound state and a resonance is nearly independent 
of the CAP strength and thus straightforward to evaluate whereas the NAC 
between two resonance states or between a bound state and a pseudocontinuum 
state is more difficult to evaluate. 
\end{abstract}

%\keywords{electronic resonances, non-Hermitian quantum chemistry, 
%non-adiabatic couplings, coupled-cluster theory, analytic-derivative theory} 
%Use showkeys class option for keyword

\maketitle

%\tableofcontents

%%%%%%%%%%%%%%%%%%%%%%%%%%%%
% Introduction              
%%%%%%%%%%%%%%%%%%%%%%%%%%%%

\section{Introduction}\label{sec:intro}
The Born-Oppenheimer (BO) separation of nuclear and electronic motion 
is a cornerstone of quantum chemistry and molecular physics of bound 
states and gives rise to the concept of electronic states and potential 
energy surfaces (PES).\cite{Born1927,Koeppel1984,Baer2002,Yarkony2012,
Curchod2018,Matsika2021,Reimers2015} It relies on the fact that the 
nuclei are significantly heavier and move slower than electrons. In 
many circumstances, one can invoke an adiabatic approximation, often referred to as the BO approximation, in which the couplings 
between different electronic states are neglected and nuclear motions 
on each PES are entirely independent from each other. 
This means that the electrons follow the nuclei instantly and the 
electronic state never changes. The BO approximation is in particular 
appropriate for the ground electronic state, which is usually well 
separated from the electronically excited states so that non-adiabatic 
interactions are highly improbable. 

%This is not quite right -- it is not just the gap, it is also the magnitude of the coupling -- i.e., how fast the wfn changes as nucleai move. But I would leave it as is for now.

% when the energetic separation between two adiabatic electronic states 
% is so big that it cannot be overcome by the energy associated with 
% nuclear motion and the electronic wave functions change only slowly 
% as the nuclei move. This applies

However, the BO approximation breaks down when two electronic states 
become close to each other, for example, near a conical intersection. 
Here, states are coupled by the kinetic energy operator, which induces 
transitions between them. Physically, this coupling results from the 
dependence of the electronic wave functions on the nuclear coordinates. Mathematically, the coupling comprises two terms:\cite{Matsika2021,
Reimers2015,Meek2016,Meek2016a} One term is a vector coupling 
involving the first derivative of the electronic wave functions, which is 
termed the non-adiabatic coupling (NAC) vector, whereas the other term is 
a scalar coupling involving the second derivative of the electronic wave 
functions. This second term is related to the diagonal Born-Oppenheimer 
correction.

% Mathematically, the coupling is given by first-derivative coupling, so-called non-adiabatic coupling (NAC), and  the second-derivative coupling. Unlike NAC, second-order derivative coupling contributes to the BO energy correction as an additional energy shift and it shows discontinuous behavior\cite{Meek2016} at conical intersections. Additionally, it has been shown\cite{Meek2016a} that whether or not we take this into account, it has little impact on the outcome.

NAC is thus a central quantity needed to describe non-adiabatic 
interactions between electronic states mediated by nuclear motion. 
These interactions give rise to non-adiabatic transitions, intensity 
borrowing, and vibronic effects. Non-adiabatic transitions result in 
radiationless relaxation, which is important in photochemistry.\cite{
Koeppel1984,Baer2002,Yarkony2012,Curchod2018,Matsika2021} 

For bound electronic states that are stable with respect to electron 
loss the computation of NACs is possible with a variety of quantum-chemical 
methods including time-dependent density functional theory,\cite{Tapavicza2013,
Ou2014,Zhang2014,Zhang2015} multiconfigurational self-consistent field 
approaches,\cite{Lengsfield1984,Galvan2016} multireference configuration 
interaction,\cite{Lengsfield1984,Saxe1985,Lischka2004} and equation-of-motion 
coupled-cluster theory.\cite{Christiansen1999,Ichino2009,
Tajti2009,Faraji2018,Koch2023} 

The situation is different for electronic resonances that decay through 
autodetachment. There is, however, substantial spectroscopic evidence 
that NAC plays a critical role for these metastable states as well. 
Recent examples include anionic resonances in pyrrole,\cite{Mukherjee2022,
Juraj:Pyrrole:2022} para-benzoquinone,\cite{West2014,Mensa2019} 
chlorobenzene,\cite{Nag2021} and hexachlorobenzene.\cite{Kumar2022} 
The evaluation of NACs with quantum-chemical methods for such systems 
has so far only been possible by approximating resonances as bound 
states and neglecting their decay.

However, this is not a good approximation because resonances are 
coupled to the continuum and cannot be associated with discrete 
states in Hermitian quantum mechanics.\cite{Moiseyev2011,Jagau2017,
Jagau2022} Rather, they manifest themselves through an increased 
density of continuum states. In this framework, 
a treatment of resonances in terms of scattering theory is 
possible.\cite{Bardsley1968,Taylor1972} In contrast, in non-Hermitian 
quantum mechanics it is possible to describe a resonance by a discrete 
state with complex energy whose real part is interpreted similarly to 
the energy of a bound state while the imaginary part corresponds to 
the decay width, that is, the inverse of the state's lifetime. This 
result can be reached using two different approaches---the formalism 
by Gamov and Siegert,\cite{Gamov1928,Siegert1939} where the resonance 
is an adiabatic state, and the formalism by Feshbach and Fano,\cite{
Feshbach1958,Fano1961,Feshbach1962,Domcke1991} where the resonance is 
treated as a bound state diabatically coupled to the continuum. In the 
former case, the decay width is a local function of the nuclear 
coordinates, whereas in the latter case it is a non-local quantity.

Nuclear motion on resonance PESs and NAC between electronic resonances 
have been analyzed by several authors.\cite{OMalley1966,Domcke1977,
Hazi1983,Bieniek1983,Domcke1984,Estrada1986,Cederbaum2003,Royal2004,
Feuerbacher2004,Haxton2007a,Haxton2007b,Benda2018a,Gyamfi2022} In 
particular, a linear vibronic coupling model for resonance-resonance 
interaction has been devised.\cite{Estrada1986} Importantly, the BO 
approximation cannot be applied to resonances without modifications 
because there is always exact energetic degeneracy between a resonance 
state and the continuum in which it is embedded. Most of the available 
theoretical treatments are based on the Feshbach formalism and 
scattering theory. This approach leads to energy-dependent non-local 
terms in the resonance potential energy surfaces as well as in the 
coupling matrix elements.\cite{Cederbaum1981,Domcke1984,Estrada1986,
Royal2004} The non-local treatment is highly accurate but its high 
computational cost and the intricate nature of the coupling matrix 
elements make it difficult to apply to more than very few nuclear 
degrees of freedom. Resonances in polyatomic molecules can thus 
only be treated within models with reduced dimensionality. 

A significant simplification is achieved by the local approximation, 
which results in an effective complex-valued potential for the motion 
of the nuclei called the boomerang model.\cite{Herzenberg1968,
Birtwistle1971,Dube1975,Dube1979}. Coupling between resonances can 
be treated within this model as well.\cite{Royal2004} However, a 
general problem of the Feshbach formalism is the definition of the 
resonance state. Especially for molecular temporary anions, there 
is no easy way to separate the resonance from the embedding continuum. 
Consequently, the boomerang model and treatments of NAC based on 
it have not been combined with state-of-the-art quantum chemistry 
methods so far. 

It thus appears worthwhile to use the Siegert formalism for the 
treatment of NACs involving resonance states. Within this approach, 
temporary anions of polyatomic molecules can be treated more easily 
using quantum chemistry methods. For example, crossing seams between 
anionic resonance states have been computed and analyzed using Siegert 
energies.\cite{Feuerbacher2004,Benda2018a}

%These issues are overcome by the former approximation, known as the "boomerang" model This model based on the assumption that the incoming electron is trapped by the  quasistationary\cite{Dube1979} electronic state $\Phi(r,R)$ and the potential for the nuclear motion associated with this state,
%\begin{align*}
%V_{local}(R) = E(R) - \frac{i}{2}\Gamma(R) 
%\end{align*}
%is local complex potential. The state has a finite decay ($\Gamma$) rate that is governed by the position of the nuclei (R). Unlike the non-local approximation, the "entry amplitude" for the incident electron that hits the target molecule is independent of the incident electron's energy. When the electron decays, the $\Phi(r,R)$ decreases, and at very large distances, it transforms into an outgoing wave. In the local approximation, the "exit amplitude" is likewise independent of the energy of the scattered electron.

%combined with either local\cite{Cederbaum1981,Hazi1983,OMalley1966} or non-local potential approximations. The later approximation provides a highly accurate description of coupling effects in resonance states, but its high computational cost and intricate nature of coupling matrix element make it difficult to apply to polyatomic systems.

In this article, we present the theory for the analytic evaluation 
of NACs within the complex absorbing potential equation-of-motion 
coupled-cluster framework (CAP-EOM-CC).\cite{Ghosh2012,Jagau2013,
Zuev2014} Our approach is based on analytic gradient theory for CAP 
methods\cite{Benda2017,Benda2018b}; the NAC elements are computed 
following the summed-state approach of Tajti and Szalay.\cite{Tajti2009} 
An important advantage of CAP-EOM-CC for the study of anionic resonances 
is that the ground and excited states of the neutral molecule and the 
bound and resonance states of the corresponding anion are treated on 
same footing. CAP-EOM-CC analytic gradients have been used to compute 
adiabatic electron affinities of temporary anions,\cite{Benda2018b} 
to locate minimum-energy crossing points between anionic and neutral 
PESs,\cite{Benda2018c} to characterize exceptional points,\cite{Benda2018a} 
and most recently to investigate the electron-energy loss spectrum 
of pyrrole.\cite{Mukherjee2022}

The article is organized as follows. In \cref{sec:theory} we present the 
theory of NACs involving resonance states and their evaluation within 
CAP-EOM-CC theory and in \cref{sec:imp} we describe our implementation. 
To illustrate complex-valued NACs obtained from these calculations, we 
use anionic states of fumaronitrile as an example. The computational 
details are outlined in \cref{sec:compd} and the results are discussed 
in \cref{sec:results}. Our concluding remarks are given in 
\cref{sec:conclusions}.

%%%%%%%%%%%%%%%%%%%%%%%%%%%%
% Theory
%%%%%%%%%%%%%%%%%%%%%%%%%%%%

\section{Theory}\label{sec:theory}
\subsection{Non-adiabatic coupling between bound electronic states} 
\label{sec:th1}

We begin with a brief description of NAC between bound electronic states 
within the context of the BO separation of variables.\cite{Matsika2021} 
The time-independent Schr\"odinger equation of a molecule can be written as
\begin{equation} \label{eq:tise}
\hat{H} \, | \Psi (r,R) \rangle = \mathcal{E} \, | \Psi (r,R) \rangle 
\end{equation}
where $R$ and $r$ denote nuclear and electronic coordinates, 
respectively. The total Hamiltonian $\hat{H}(r,R) = \hat{T}_n(R) + 
\hat{H}^{el}(r,R)$ consists of the electronic Hamiltonian $\hat{H}^{el}$ 
and the nuclear kinetic energy operator $\hat{T}_n$. The total wave function 
$\Psi_T (r,R)$ for a vibronic state $T$ can be expressed using a sum over 
products of nuclear and electronic wave functions,
\begin{equation} \label{eq:vibro}
| \Psi_T (r,R) \rangle = \sum_I | \Phi_I (r;R) \; 
\mathbf{\xi}^T_I (R) \rangle~.
\end{equation}
Within the BO separation, the electronic wave functions 
$|\Phi_I (r;R) \rangle$ are obtained by solving the 
electronic Schr\"odinger equation at fixed nuclear positions, 
\begin{equation} \label{eq:tiseel}
\hat{H}^{el} \; | \Phi_I (r;R) \rangle = E_I(R) \; | 
\Phi_I (r;R) \rangle 
\end{equation}
with $E_I$ as electronic energy of adiabatic state $I$.  

The equations that determine the nuclear wave functions $\mathbf{\xi}^T_I(R)$ 
are obtained by plugging \cref{eq:vibro} into \cref{eq:tise} 
\begin{equation} \label{eq:nucwfn1}
\Big[ \hat{T}_n(R) + \hat{H}^{el}(r,R) \Big] \sum_I |\Phi_I (r;R) \, 
\mathbf{\xi}^T_I(R) \rangle = \mathcal{E}^T \sum_I |\Phi_I (r;R) \, 
\mathbf{\xi}^T_I(R) \rangle 
\end{equation}
and projection of the Schr\"odinger equation onto an electronic state 
$\langle \Phi_J |$. This yields
\begin{equation} \label{eq:nucwfn2}
\Big[ \hat{T}_n + E_J(R) \Big] | \, \mathbf{\xi}^T_J(R) \rangle 
+ \sum_I \hat{\Lambda}_{JI} | \, \mathbf{\xi}^T_I(R) \rangle 
= \mathcal{E}^T | \, \mathbf{\xi}^T_J(R) \rangle~.
\end{equation}
The first term in \cref{eq:nucwfn2} describes nuclear motion on an 
isolated PES $E_J(R)$, while $\hat{\Lambda}_{JI}$ ($I\neq J$) is 
the NAC between states $J$ and $I$. Within the BO approximation, 
$\hat{\Lambda}_{JI}$ is neglected resulting in the separation of 
electronic and nuclear degrees of freedom. However, to describe 
non-adiabatic transitions between electronic states, this term 
needs to be considered. The coupling matrix element is given by
\begin{equation} \label{eq:nacFG}
\hat{\Lambda}_{JI} = \sum_\alpha \frac{1}{2M_\alpha} \big[ 2 \, 
\mathcal{F}_{JI}^\alpha (R) \cdot \nabla_\alpha + 
\widetilde{\mathcal{F}}_{JI}^\alpha (R) \big] 
\end{equation}
where the sum runs over all nuclear indices $\alpha$ 
and $M_\alpha$ and $\nabla_\alpha$ denote the corresponding masses 
and gradients. The latter are usually evaluated in Cartesian 
coordinates. The contributions $\mathcal{F}_{JI}$ and 
$\widetilde{\mathcal{F}}_{JI}$ are given as  
\begin{align} \label{eq:nacF}
\mathcal{F}_{JI}^\alpha (R) &= \langle \Phi_J(r;R) 
| \nabla_\alpha | \Phi_I (r;R) \rangle~, \\
\widetilde{\mathcal{F}}_{JI}^\alpha (R) &= 
\langle \Phi_J (r;R) | \nabla_\alpha^2 | 
\Phi_I (r;R) \rangle = \nabla_\alpha \mathcal{F}_{JI}^\alpha (R) 
+ \sum_K \mathcal{F}_{JK}^\alpha (R) \, \mathcal{F}_{KI}^\alpha (R)~.
\label{eq:nacG} \end{align}
The last identity shows that $\mathcal{F}_{JI}^\alpha (R)$ is 
sufficient to describe the coupling.\cite{Baer2002} 
Note, however, that it only holds if the electronic wave functions 
form a complete basis. 

Semiclassical treatments of nuclear motion\cite{Matsika2021} 
commonly neglect $\widetilde{\mathcal{F}}_{JI}^\alpha (R)$ and consider 
only $\mathcal{F}_{JI}^\alpha (R)$, which is called derivative coupling 
or NAC. Although it has been argued that the other term, $\widetilde{
\mathcal{F}}_{JI}^\alpha (R)$, in general cannot be neglected\cite{
Reimers2015,Meek2016}, it has also been shown how to account for it 
in the simulation of non-adiabatic dynamics.\cite{Meek2016a} The impact 
of $\widetilde{\mathcal{F}}_{JI}^\alpha (R)$ on the final results appears 
to be limited.

%There is evidence\add{\cite{Meek2016a}} that 
% \textcolor{red}{$\widetilde{\mathcal{F}}_{JI}^\alpha (R)$} has little 
% impact on non-adiabatic dynamics and 

%The $\mathcal{G}_{IJ}$ term is simplified further in terms of 
% $\mathcal{F}_{IJ}^\alpha(R)$ by using the following,
%\begin{align}
%      \langle \nabla_{R_\alpha} \Phi_I | \Phi_K \rangle = - \langle \Phi_I |  \nabla_{R_\alpha} \Phi_K %\rangle \;\; ,  \;\;  \text{since $\langle \Phi_I | \Phi_K \rangle = \delta_{IJ}$}  %\label{grad_psi_IJ} 
%\end{align}
% The off-diagonal $\mathcal{F}_{IJ}$ term leads to the coupling between the electronic states, whereas, $\mathcal{G}_{IJ}$ term contributes to the diagonal term as an additional energy shift. When the potential energy surfaces cross, $\mathcal{G}_{IJ}$ term exhibits discontinuous behavior\cite{Meek2016,Tully1971}.  However, it has been shown in the literature\cite{Meek2016a} that whether or not the $\mathcal{G}_{IJ}$ term is properly taken into account has little impact on the dynamics.

The elements of the NAC vector can be evaluated by 
noticing that for an exact solution of the electronic Schr\"odinger 
equation, the nuclear gradient of state $I$ ($\mathbf{G}_I$) satisfies 
the Hellman-Feynman theorem, 
\begin{equation}
G_I^\alpha \equiv \nabla_\alpha \langle \Phi_I | \hat{H}^{el} | 
\Phi_I \rangle = \langle \Phi_I | (\nabla_\alpha \hat{H}^{el}) | 
\Phi_I \rangle~. \label{eq:feynman1} 
\end{equation}
By generalization to off-diagonal matrix elements, 
i.e., interstate couplings, one obtains for $\mathcal{F}_{JI}^\alpha (R)$
\begin{equation} \label{eq:feynman2}
0 = \nabla_\alpha \langle \Phi_I | \hat{H}^{el} | \Phi_J \rangle 
= (E_I - E_J) \langle \Phi_I | \nabla_\alpha | \Phi_J \rangle 
+ \langle \Phi_I | (\nabla_\alpha \hat{H}^{el}) | \Phi_J \rangle~.
\end{equation}

From \cref{eq:feynman2} it follows that the derivative coupling 
can be evaluated as
\begin{equation} \label{eq:nacvec}
\mathcal{F}^\alpha_{IJ} (R) = \frac{\langle \Phi_I | (\nabla_\alpha 
\hat{H}^{el}) | \Phi_J \rangle}{E_J - E_I} 
\end{equation}
where $h^\alpha_{IJ} \equiv \langle \Phi_I | (\nabla_\alpha \hat{H}^{el}) | 
\Phi_J \rangle$ can be viewed as an interstate generalization of the 
nuclear gradient that is called the NAC force.\cite{Faraji2018,Matsika2021}
\cref{eq:nacvec} illustrates that the derivative coupling becomes 
large when two potential energy surfaces are nearly degenerate. 
Since $\mathcal{F}_{JI}^\alpha (R)$ enters the 
Schr\"odinger equation for the nuclei, Eq. \eqref{eq:nucwfn2}, through 
the scalar product with the nuclear velocity in Eq. \eqref{eq:nacFG}, 
non-adiabatic transitions are also more likely when the nuclei move 
fast. Conversely, the adiabatic approximation is recovered when the 
nuclei are moving infinitesimally slow. We note that the sign of the 
vectors $h_{IJ}$ and $\mathcal{F}_{IJ}$ is arbitrary: A change of phase 
in either $\Phi_I$ or $\Phi_J$ induces a sign change in $h_{IJ}$ and 
$\mathcal{F}_{IJ}$ but the resulting wave function is still a solution 
to \cref{eq:tiseel}. Also, the elements of $h_{IJ}$ and $\mathcal{F}_{IJ}$ 
sum up to zero because of translational invariance.\cite{
Tommasini2001,Fatehi2011}

%%%%%%%%%%%%%%%%%%%%%%%%%%%%%%%%%%%%%%%%%%%%%%%%%%%%%%%%%%%%%%%%%%%%%%%%%

\subsection{Non-adiabatic coupling between resonances based on complex 
absorbing potentials} \label{sec:th2}

The practitioners of bound-state quantum chemistry need not be concerned 
that the $|\Phi_I(r;R) \rangle$ in Eq. \eqref{eq:vibro} 
only form a complete basis if continuum states are included as they work 
with a basis-set representation of finite size. However, this needs to 
be reconsidered when dealing with NACs where one or both of the coupled 
states is an electronic resonance because these states are embedded 
in the continuum.\cite{Moiseyev2011,Jagau2017,Jagau2022} 

Although it is possible to extend the BO ansatz to Hilbert spaces of 
infinite dimensions,\cite{Baer2002} we will follow the conventional 
approach here and work with a finite set of electronic functions. 
We accomplish this in the framework of non-Hermitian 
quantum chemistry where electronic resonances are described as discrete 
states separated from continuum. In contrast, in Hermitian quantum 
chemistry, resonances are not discrete states but correspond to an 
increased density of continuum states.

Our computational treatment of NAC employs the Siegert representation 
of electronic resonances.\cite{Siegert1939} This means that we consider 
eigenstates of the electronic Hamiltonian that diverge exponentially 
in space and have complex energy
\begin{equation} \label{eq:eres}
E_\text{res} = E - i \, \Gamma/2~. 
\end{equation}
The real part of the energy corresponds to the position of the resonance, 
whereas the imaginary part is related to the decay width $\Gamma$. For 
bound states, $\Gamma = 0$. The Siegert states can be included in the 
manifold of electronic wave functions $| \Phi_I (r;R) 
\rangle$ in Eq. \eqref{eq:vibro} in a straightforward manner. $E_\text{res}$ 
from Eq. \eqref{eq:eres} then depends parametrically on the nuclear 
coordinates and the resonance width $\Gamma$ is a local quantity.  

Alternatively, Eq. \eqref{eq:eres} can be obtained following the theory 
by Feshbach and Fano.\cite{Feshbach1958,Fano1961,Feshbach1962} Importantly, 
$\Gamma$ is a non-local quantity in this framework. We discuss the 
treatment of non-adiabatic effects based on this latter approach in 
Sec. \ref{sec:th3}. Also, we note that the coupling between diabatic 
resonance states and the ensuing nuclear dynamics have been analyzed 
in Ref. \citenum{Estrada1986}.  

Siegert states and energies can be computed using different techniques, 
in particular using complex scaling\cite{Aguilar1971,Balslev1971} or, 
alternatively, complex absorbing potentials (CAPs).\cite{Jolicard1985,
Riss1993} Here, we use the CAP method where an imaginary potential 
$\hat{W}(r)$ is added to the electronic Hamiltonian $\hat{H}^\text{el}$ 
according to 
\begin{equation} \label{eq:hcap}
\hat{H}^\text{el} (\eta) = \hat{H}^\text{el} - i \, \eta \, \hat{W}(r)~. 
\end{equation}
Different functional form have been suggested for $\hat{W}$, all 
of which bring the diverging resonance wave function into an 
$\mathcal{L}^2$-integrable form. Here we use a shifted quadratic 
potential defined as
\begin{equation}
\label{eq:W_cap}
\begin{aligned}
\hat{W} (r) &= \hat{W}_x (x) + \hat{W}_y (y) + \hat{W}_z (z) \\
\hat{W}_{\alpha}(\alpha) &=
\begin{cases}
0 & \text{if $\left| \alpha - o_\alpha \right| \leq r_\alpha^0 $} \\
\text{$\left( \left| \alpha - o_\alpha \right| - r_\alpha^0 \right)^2$} & 
\text{otherwise} \end{cases}
\; , \quad \alpha = (x,  y,  z)
\end{aligned}
\end{equation}
where $( r_{x}^0, \, r_{y}^0, \, r_{z}^0 )$ and $( o_{x}^0, \, o_{y}^0, 
\, o_{z}^0 )$ are the onset and origin in each dimension and $\eta$ is 
a strength parameter. 

The CAP Hamiltonian $\hat{H}^\text{el}(\eta)$ has complex eigenvalues 
and it can be shown that some of them converge to true Siegert 
energies defined according to Eq. \eqref{eq:eres} in the limit 
$\eta \to 0^+$.\cite{Riss1993} However, in a finite basis set where 
$\hat{H}^\text{el}(\eta)$ is represented approximately, this limit 
is not meaningful. In this work, we follow the most common strategy 
and determine an optimal CAP strength $\eta_\text{opt}$ by minimizing 
the perturbation of the energy in first order, which yields the 
criterion $\text{min} |\eta \, dE / d\eta |$.\cite{Riss1993} The 
remaining eigenvalues of $\hat{H}(\eta)$, which do not correspond to 
resonances, are either bound states with real energy or pseudocontinuum 
states whose complex energy changes much more rapidly with $\eta$ than 
that of resonances so that no $\eta_\text{opt}$ can be determined. 

To derive equations for the nuclear motion, we reconsider Eqs. 
\eqref{eq:tise} and \eqref{eq:vibro} and assume that some electronic 
wave functions $| \Phi_I (r;R) \rangle$ have complex 
energy. Eqs. \eqref{eq:nucwfn1} and \eqref{eq:nucwfn2} do not formally 
change, but the potential energy surface $E_J(R)$ is complex-valued if 
one projects onto a resonance state $\langle \Phi_J(r;R) 
|$. If $\langle \Phi_J (r;R) |$ is a bound state, 
$E_J(R)$ is real-valued whereas $\hat{\Lambda}_{JI}$ is complex-valued 
for all $I$ that correspond to resonance states. 

The definitions of the coupling matrix elements, Eqs. 
\eqref{eq:nacFG}--\eqref{eq:nacG}, do not change when one or both of 
the coupled states have complex energy. Also, it is still possible 
to evaluate the derivative coupling from the NAC force through Eq. 
\eqref{eq:nacvec}. This shows that the derivative coupling diverges 
at exceptional points where the real and imaginary parts of two 
resonance energies $E_I$ and $E_J$ become identical. An important 
difference between real-valued and complex-valued NAC is also apparent 
from Eq. \eqref{eq:nacvec}: The vectors $\mathcal{F}_{IJ}$ and $h_{IJ}$ 
are necessarily collinear if $E_I$ and $E_J$ are both real but this 
is not the case if one or both of them are complex. Rather, the four 
vectors $\text{Re}(\mathcal{F}_{IJ})$, $\text{Im}(\mathcal{F}_{IJ})$, 
$\text{Re}(h_{IJ})$, and $\text{Im}(h_{IJ})$ can all point in different 
directions. 

%%%%%%%%%%%%%%%%%%%%%%%%%%%%%%%%%%%%%%%%%%%%%%%%%%%%%%%%%%%%%%%%%%%%%%%
\subsection{Non-adiabatic coupling between resonances based on 
Feshbach's projection formalism}
\label{sec:th3}

As an alternative to CAPs, Feshbach's projection formalism can 
be used to define resonance states. In the following, we use the 
ansatz by O'Malley,\cite{OMalley1966} later extended by Royal 
\textit{et al.},\cite{Royal2004} to obtain further insights into 
NAC between resonance states and the meaning of complex-valued 
NAC vectors between Siegert states obtained from CAP calculations. 

In the Feshbach formalism,\cite{Feshbach1958,Fano1961,Feshbach1962,
Domcke1991} the total wave function $| \Psi(r,R) \rangle$ from Eq. 
\eqref{eq:tise} is expressed as 
\begin{equation} \label{eq:psi_PQ}
| \Psi \rangle = \hat{Q} | \Psi \rangle + \hat{P} | \Psi \rangle 
= | \Psi_Q \rangle + | \Psi_P \rangle 
\end{equation}
where the projectors $\hat{Q}$ and $\hat{P}$ operate on the 
electronic part of the total wave function and are defined as 
\begin{align} \label{eq:Q_proj}
\hat{Q} &= \sum_I^n | \Phi_I (r;R) \rangle \, 
\langle \Phi_I (r;R) |~, \\
\hat{P} &= 1 - \hat{Q} = \int_0^\infty dE |\chi_E \rangle \langle \chi_E | 
\label{eq:P_proj}
\end{align}
with $|\Phi_I (r;R) \rangle$ as discrete $L^2$-normalized 
electronic resonance states and $| \chi_E \rangle$ as $\delta$-normalized 
scattering states. Using Eq. \eqref{eq:vibro} one obtains for 
$| \Psi_Q \rangle$ the explicit form
\begin{equation} \label{eq:Psi_Q}
|\Psi_Q (r,R) \rangle = \sum_I |\Phi_I (r;R) \, \xi_I (R) \rangle~.
\end{equation}
Applying the Hamiltonian to Eq. \eqref{eq:psi_PQ} yields the following 
coupled equations: 
\begin{align} \label{eq:P_sol}
\hat{P} (\hat{H} -\mathcal{E}) \, \hat{P} | \Psi \rangle &= 
-\hat{P} \hat{H} \hat{Q} \, | \Psi \rangle~, \\
\hat{Q} (\hat{H} - \mathcal{E}) \, \hat{Q} | \Psi \rangle &= 
-\hat{Q} \hat{H} \hat{P} \, |\Psi \rangle~. \label{eq:Q_sol}	 
\end{align}
By plugging Eq. \eqref{eq:P_sol} into Eq. \eqref{eq:Q_sol}, a 
projected Schr\"odinger equation is obtained for the discrete 
states, which reads 
\begin{equation} \label{eq:PQ_proj}
\hat{Q}\hat{H}\hat{Q} \, | \Psi_Q \rangle + \Big[ 
\hat{Q}\hat{H}\hat{P} \, (\mathcal{E} - \hat{P}\hat{H}\hat{P})^{-1} \, 
\hat{P}\hat{H}\hat{Q} \Big] \, | \Psi_Q \rangle = \mathcal{E} \, 
|\Psi_Q \rangle~.
\end{equation} 
Here, the projected Hamiltonian $\hat{Q}\hat{H}\hat{Q}$ is corrected 
by a complex level-shift operator $\mathcal{S}$ defined by
\begin{align}
\mathcal{S}& = \hat{Q}\hat{H}\hat{P} \, (\mathcal{E} - \hat{P}\hat{H}\hat{P})^{-1} 
\, \hat{P}\hat{H}\hat{Q} = \hat{Q}\hat{H}\hat{P} \; G_P \; 
\hat{P}\hat{H}\hat{Q} \nonumber \\
&= \sum_{IJ}^n |\Phi_I (r;R) \rangle 
\langle \Phi_I (r;R) | \, 
\hat{H}^{el} P \, G_P \, P\hat{H}^{el} \, |\Phi_J (r;R) \rangle 
\langle \Phi_J (r;R) | \nonumber \\
& = \sum_{IJ}^{n} |\Phi_I(r;R)\rangle \, 
\mathcal{S}_{IJ} \, \langle \Phi_J(r;R) | \label{eq:lshift} 
\end{align}
with $G_P = \lim_{\epsilon \to 0} (\mathcal{E} - \hat{P}\hat{H}\hat{P} 
+ i\epsilon)$ as Green's function in the $P$-space. 

By integrating over the electronic coordinates, one obtains from 
Eq. \eqref{eq:PQ_proj} the following equation for the nuclear wave 
function\cite{OMalley1966,Domcke1984,Royal2004} 
\begin{equation}
\Big[ \hat{T}_n + E_J(R) + \mathcal{S}_{JJ} (R) \Big] |\xi_J (R) \rangle 
+ \sum_I \Big[ \hat{\Lambda}_{JI} + \mathcal{S}_{JI} \Big] \, |\xi_I(R) \rangle 
= \mathcal{E} \, |\xi_J (R) \rangle~. 
\label{eq:nucres}
\end{equation}
Eq. \eqref{eq:nucres} governs nuclear motion in the resonance state 
and is the equivalent of Eq. \eqref{eq:nucwfn2}. The first term 
describes the motion on an isolated resonance PES while the second 
term is the NAC. Evidently, both terms include energy-dependent and 
non-local contributions due to $\mathcal{S}$ that are absent in 
Eq. \eqref{eq:nucwfn2}.

It is, however, often possible to invoke a local complex potential 
approximation, which results in the boomerang model.\cite{Herzenberg1968,
Birtwistle1971,Dube1975} This yields for the elements of $\mathcal{S}$
\begin{align}
\mathcal{S}_{II} &= \Delta_I - i \, \Gamma_I/2  \quad \text{for}~I=J~, 
\label{eq:boom1} \\
\mathcal{S}_{IJ} &= - i \, \sqrt{\Gamma_I \, \Gamma_J}\, /2 \quad \text{for}~I\neq J ~, 
\label{eq:boom2}
\end{align}
where $\Delta_I$ represents an energy shift, which is set to zero in 
the coupling term $\mathcal{S}_{IJ}$. Evidently, the coupling term 
$\mathcal{S}_{IJ}$ vanishes if one of the coupled states is bound 
($\Gamma=0$). Likewise, the diagonal term $\mathcal{S}_{II}$ vanishes 
for bound states as well. 

Eqs. \eqref{eq:nucres}--\eqref{eq:boom2} provide a basis for the 
interpretation of NACs obtained from CAP calculations. In the Feshbach 
formalism, $\hat{\Lambda}_{JI}$ describes the NAC between two discrete 
states and the resonance character comes about solely due to 
$\mathcal{S}_{IJ}$. This suggests to associate $\text{Im}(\mathcal{F}_{IJ})$ 
obtained in a CAP calculation with $\mathcal{S}_{IJ}$ 
and to interpret $\text{Re}(\mathcal{F}_{IJ})$ as an analog of the 
NAC between bound states. However, this comparison is problematic for 
at least two reasons: First, $\mathcal{S}_{IJ}$ does not depend on 
the nuclear coordinates, meaning it should provide 
the same contribution to all elements of the NAC vector, whereas 
$\text{Im}(\mathcal{F}_{IJ})$ inherently depends on the nuclei so 
that all elements of the vector are different. Second, all results 
obtained in a CAP calculation are subject to an unwanted dependence 
on the CAP strength $\eta$. 

We note that recent experiments\cite{Juraj:Pyrrole:2022} provide a 
concrete illustration that non-adiabatic transitions between resonances 
are modulated by specific vibrations, which can only be explained by 
the dependence of NACs on nuclear coordinates. However, 
these observations do not rule out that only the real part of the NAC 
vector depends on the nuclear coordinates and the imaginary part is 
coordinate independent.

\subsection{Evaluation of non-adiabatic couplings within CAP-EOM-CC 
framework} 
\label{sec:NACEOM}

The computation of NACs between bound states within the EOM-CC 
framework has been discussed in Refs. \citenum{
Christiansen1999,Tajti2009,Ichino2009,Faraji2018,Koch2023}. 
Notably, the different formulations are not numerically identical.
The recent the work of Kj{\o}nstad and Koch\cite{Koch2023} as well 
as the earlier work by Christiansen\cite{Christiansen1999} employ 
CC response theory. In Ref. \citenum{Koch2023}, the NAC is obtained 
from a biorthonormal formulation in which only the right state 
is differentiated. In contrast, the works by Ichino \textit{et 
al.}\cite{Ichino2009}, by Tajti and Szalay,\cite{Tajti2009} and 
by Faraji \textit{et al.}\cite{Faraji2018} use CC gradient theory.

Here, we use the second strategy, more specifically, 
the summed-state approach introduced by Tajti and Szalay,\cite{
Tajti2009} and combine it with analytic-gradient theory for CAP 
methods.\cite{Benda2017} This allows us to compute NACs between 
CAP-EOM-CCSD states. 

In EOM-CC theory,\cite{Krylov2008,Bartlett2012,Sekino1984,Stanton1993,
Stanton1994,Nooijen1995} target states $| \Phi_I \rangle$, 
$| \Phi_J \rangle$, \dots are defined by applying 
configuration interaction (CI) like linear excitation operators $R_I$, 
$R_J$, \dots to the CC reference state 
\begin{equation} \label{eq:eomR}
| \Phi_I \rangle = R_I | \, \Phi_\text{ref} 
\rangle = R_I \, e^T \, | 0 \rangle
\end{equation}
where $| 0 \rangle$ is the reference determinant, usually the 
Hartree-Fock (HF) determinant, and $T$ stands for the coupled-cluster 
amplitudes that satisfy the CC equations for the reference state 
$| \Phi_\text{ref} \rangle$. The left EOM-CC states 
$\langle \Phi_I |$, $\langle \Phi_J |$, 
\dots are not the conjugates of the right EOM-CC states but rather 
chosen as
\begin{equation} \label{eq:eomL}
\langle \Phi_I | = \langle 0 | \, L_I^\dagger \, e^{-T}
\end{equation}
where $L_I$ is a CI-like excitation operator as well. The EOM-CC 
energies and eigenvectors are obtained by solving the eigenvalue 
equations
\begin{align}
\bar{H} \, R_I \, |0 \rangle &= E_I \, R_I \, | 0 \rangle~, \label{eq:eomseR} \\
\langle 0 | \, L_J^\dagger \, \bar{H} &= E_J \, \langle 0 | \, L_J^\dagger~, 
\label{eq:eomseL} \\
\langle 0 | L_J^\dagger R_I | 0 \rangle &= \delta_{IJ}~. \label{eq:biorth}
\end{align}
where $\bar{H} = e^{-T} \hat{H} e^T$ is the similarity-transformed 
Hamiltonian.

Depending on the choice of $R$ and $L$, \cref{eq:eomR} and \cref{eq:eomL} 
describe excited, electron-attached, or ionized states. The truncation 
level of $T$, $R$, and $L$ defines the EOM-CCSD, EOM-CCSDT, and so 
forth models. In the context of this work, where we focus on bound 
and temporary radical anions, we consider the EOM-EA-CCSD method where 
$R$ and $L$ are electron-attaching operators comprising one-particle 
($1p$) and two-particle-one-hole ($2p1h$) excitations.\cite{Nooijen1995} 

To evaluate the NAC between two EOM-CCSD states $I$ and $J$, we consider 
an artificial summed state $| \Phi_{I+J} \rangle \equiv 
| \Phi_I \rangle + | \Phi_J \rangle$ 
and its gradient $\mathbf{G}_{I+J}$, which is related to the NAC force 
$\mathbf{h}_{IJ}$ according to\cite{Tajti2009}
\begin{align}
G^\alpha_{I+J} &= \langle \Phi_{I+J} | (\nabla_\alpha 
\bar{H}) | \Phi_{I+J} \rangle = \langle \Phi_I 
| (\nabla_\alpha \bar{H}) | \Phi_I \rangle + \langle 
\Phi_J | (\nabla_\alpha \bar{H}) | \Phi_J 
\rangle + 2 \langle \Phi_I | (\nabla_\alpha \bar{H}) 
| \Phi_J  \rangle \nonumber \\
&= G^\alpha_I + G^\alpha_J + 2 \langle \Phi_I | 
(\nabla_\alpha \bar{H}) | \Phi_J  \rangle = G^\alpha_I 
+ G^\alpha_J + 2 \, h^\alpha_{IJ}~.
\label{eq:gradij}
\end{align}
From \cref{eq:gradij} it follows that the NAC force can be evaluated as 
\begin{equation}
h^\alpha_{IJ} = 0.5 \, ( G^\alpha_{I+J} - G^\alpha_I - G^\alpha_J)
\label{eq:nacforce} 
\end{equation}
where the summed-state gradient vector is computed in analogy to the 
proper gradient vectors $\mathbf{G}_I$ and $\mathbf{G}_J$. The theory 
of analytic EOM-CC gradients\cite{Stanton1993b} is based on the general 
theory of molecular property calculations in CC theory. For an efficient 
implementation, the gradient is expressed in terms of differentiated 
matrix elements over atomic-orbital integrals. A generic expression is 
\begin{equation} \label{eq:ccgrad}
G^\alpha_I = \sum_{\mu\nu} \gamma_{\mu\nu} \, \mathcal{H}^\alpha_{\mu\nu} + 
\sum_{\mu\nu\rho\sigma} \, \Gamma_{\mu\nu\rho\sigma} \langle \mu\nu || 
\rho\sigma \rangle^\alpha + \sum_{\mu\nu} I_{\mu\nu} \, S^\alpha_{\mu\nu}
\end{equation}
where $\boldsymbol{\gamma}$, $\boldsymbol{\Gamma}$, and $\mathbf{I}$ 
are density matrices, whose exact definitions depend on the EOM-CC 
model, and $\mathcal{H}^\alpha$, $\langle \mu \nu || \rho \sigma 
\rangle^\alpha$, and $S^\alpha$ are derivatives of the one-electron, 
two-electron, and overlap integrals. 

Because $\bar{H}$ is not symmetric, 
\begin{equation}
h^{\alpha}_{IJ} = \langle \Phi_I | (\nabla_\alpha \bar{H}) 
| \Phi_J \rangle \neq \langle \Phi_J 
| (\nabla_\alpha \bar{H}) | \Phi_I \rangle = h^\alpha_{JI} 
\label{eq:naceom}
\end{equation}
similar to other interstate properties in EOM-CC theory.\cite{Stanton1993}
Possible solutions are to consider either the geometric\cite{Ichino2009,
Faraji2018} or the arithmetic mean,\cite{Tajti2009} here we choose the 
former approach. 

%Unlike $I, J$ states, the Hellman-Feynman theorem \cref{Hell_Feynman} 
%does not hold for $\Phi_{I+J}$ state because the summed state is not 
%an eigenstate. If we define the Hellman-Feynman gradient of $\Phi_{I+J}$ state,
%the violation arises in the third term, which is interpreted as the NAC force.
% the NAC force in the CAP-EOM-CC framework, analogous to the CI formalism\cite{Fatehi2011,Zhang2014}, % can be expressed as a non-Hermitian complex-valued extension of \cref{nac_force}.

One concern regarding the computation of the derivative coupling 
$\mathcal{F}_{IJ}$ is that \cref{eq:nacvec} does not hold for 
approximate solutions of the Schr{\"o}dinger equation. Rather, 
the result of \cref{eq:nacvec} corresponds to a modified derivative 
coupling $\mathcal{F}^C_{IJ}$ where $\nabla_\alpha$ does not 
operate on the HF wave function but only on the CC and EOM-CC 
amplitudes.\cite{Tajti2009,Faraji2018} Whereas 
$\mathcal{F}^C_{IJ}$ is translationally invariant, the full 
expression, which includes the derivative of the HF wave function, 
violates translational invariance, which is why it is commonly 
omitted.\cite{Faraji2018,Fatehi2011,Zhang2014} This is also done 
in the present work where we use \cref{eq:nacvec} to compute 
$\mathcal{F}^C_{IJ}$.

In a computation of NACs between CAP-EOM-CCSD states, the wave 
function parameters, energy eigenvalues, and gradient vectors 
become complex.\cite{Zuev2014,Jagau2013,Benda2017} However, Eqs. 
\eqref{eq:eomR}-\eqref{eq:naceom} do not formally change 
except that the usual scalar product is replaced by the 
c-product.\cite{Moiseyev2011,Moiseyev1978} We point out that 
this is only the case if the CAP is included in the Hamiltonian 
at the HF level. For projected CAP methods,\cite{Sommerfeld2001,
Ehara2012,Gayvert2022} which offer the advantage of reduced 
computational cost, a separate gradient theory would need to 
be worked out. Importantly, all integrals over atomic orbitals 
and their derivatives with respect to nuclear displacements 
are real-valued in CAP-EOM-CCSD, while the density matrices 
in \cref{eq:ccgrad} are complex-valued. Also, the derivative 
of the one-electron Hamiltonian $\mathcal{H}^\alpha$ contains 
some extra terms that result from the differentiation of the 
CAP and the dependence of the CAP origin on the nuclear 
coordinates.\cite{Benda2017}

To analyze the dependence of $h_{IJ}^\alpha = \langle 
\Phi_I | (\nabla_\alpha \bar{H}(\eta)) | 
\Phi_J \rangle$ on $\eta$, the definition of the 
CAP Hamiltonian from Eq. \eqref{eq:hcap} can be used. This yields
\begin{equation} \label{eq:cxnac}
h_{IJ}^\alpha (\eta) = \langle \Phi_I | (\nabla_\alpha 
\bar{H}) | \Phi_J \rangle - i \, \eta \, \langle 
\Phi_I | (\nabla_\alpha \bar{W}) | \Phi_J 
\rangle~, 
\end{equation}
which illustrates that $h_{IJ}^\alpha (\eta)$ depends linearly on $\eta$ 
for large $\eta$ values. We note that this is similar to the energy but 
different to molecular properties that can be formulated as expectation 
values. For the latter quantities, there is no term that depends on $\eta$ 
explicitly and the overall dependence on $\eta$ is determined entirely 
by that of the density matrix.\cite{Jagau2016} Eq. \eqref{eq:cxnac} 
suggests that $h_{IJ}^\alpha (\eta)$ can be de-perturbed in analogy to 
the energy\cite{Jagau2013} by removing the term that depends on $\eta$ 
explicitly. 

%{\bf This is a very interesting observation!  Should we then de-perturb the NAC, by removing eta term? Or there are arguments against it? If there are, should not we spell them out here?}
%%%%%%%%%%%%%%%%%%%%%%%%%%%%
% Implementation
%%%%%%%%%%%%%%%%%%%%%%%%%%%%%%%%%%%%%%%%%%%%%%%%%%%%%%%%%%%%%%%%%%%%%%%%%%
\section{Implementation}\label{sec:imp}

All expressions for evaluating the NAC force $h_{IJ}$ and the derivative 
coupling $\mathcal{F}_{IJ}$ are implemented in the Q-Chem electronic 
structure program.\cite{Epifanovsky2021} Our implementation is able to 
compute couplings between CAP-EOM-EA-CCSD states, between CAP-EOM-IP-CCSD 
states, and between CAP-EOM-EE-CCSD states. However, couplings between 
CAP-EOM-EE-CCSD states and the CAP-CCSD reference state are not implemented. 
Also, our implementation requires to include all electrons in the correlation 
treatment because the implementation of analytic CAP-EOM-CCSD 
gradients\cite{Benda2017} on which our work is based has the same restriction.

The following steps are taken to compute $h_{IJ}$ and $\mathcal{F}_{IJ}$:
\begin{enumerate}
\item Solve the CAP-HF and CAP-CCSD equations for the reference state. 
\item Solve the right and left CAP-EOM-EA-CCSD equations, \cref{eq:eomseR} 
and \cref{eq:eomseL}, for the coupled states $I$ and $J$. 
\item Solve the amplitude response and orbital response equations 
for states $I$, $J$, and $I+J$ and construct the density matrices 
$\boldsymbol{\gamma}$, $\boldsymbol{\Gamma}$, and $\mathbf{I}$ for 
each state. 
\item Evaluate the gradient vectors $\mathbf{G}_I$, $\mathbf{G}_J$, 
and $\mathbf{G}_{I+J}$ using \cref{eq:ccgrad}.
\item Compute NAC forces $h_{IJ}$ using \cref{eq:nacforce} and 
derivative couplings $\mathcal{F}_{IJ}$ using \cref{eq:nacvec}.
\end{enumerate}

To verify our implementation, we evaluated the summed-state gradient 
$\mathbf{G}_{I+J}$ through numerical differentiation. Note that 
special attention has to be paid to the relative phase of the two 
coupled states in these calculations.  

% The addition of CAP has no impact on the scaling of the EOM-EA-CCSD. 
% However, dealing with complex numbers leads the storage requirements 
% to increase by a factor of 2, and the computational costs to increase 
% by a factor of 4. The additional cost of gradient calculations is insignificant. 

%%%%%%%%%%%%%%%%%%%%%%%%%%%%
% Computational details
%%%%%%%%%%%%%%%%%%%%%%%%%%%%%%%%%%%%%%%%%%%%%%%%%%%%%%%%%%%%%%%%%%%%%%%%
\section{Computational details}\label{sec:compd}

As an illustration of complex-valued NACs between CAP-EOM-EA-CCSD 
states, we consider anionic states of fumaronitrile (trans-CN-CH=CH-CN, 
point group $C_{2h}$). We chose fumaronitrile as an 
example for the following reasons: First, electron attachment to an 
out-of-plane $\pi^*$ molecular orbital (MO) produces a bound anion 
($^2\text{B}_g$ state) with an energy lower than that of the neutral 
ground state. Second, multiple resonance states with 
different symmetry exist. Ehara and Sommerfeld reported four anionic 
resonance states,\cite{Ehara2017} of in-plane ($^2\text{A}_g$, 
$^2\text{B}_u$) and out-of-plane ($^2\text{A}_u$, $^2\text{B}_g$) 
character using the symmetry adapted cluster (SAC)-CI ansatz and 
a projected CAP.\cite{Ehara2012}

Here, we consider the bound anion ($^2\text{B}_g$) and the resonance 
states of $^2\text{A}_g$ and $^2\text{A}_u$ symmetry and compute the 
NAC between bound and pseudocontinuum states, between bound and 
resonance states, and between two resonance states. The geometry 
of neutral fumaronitrile was optimized in the $xy$-plane using 
B3LYP\cite{Becke1993,Stephens1994_b3lyp}/cc-pVTZ\cite{Dunning1989} 
and is provided in the SI. In the following, we 
refer to the central carbon atoms of fumaronitrile as C1 and C2, 
while the outer carbon atoms, which belong to the cyano groups, 
are named C3 and C4. Note that C1 and C2, C3 and C4, as well as 
the nitrogen and hydrogen atoms are pairwise equivalent due to 
symmetry.

In the CAP-EOM-EA-CCSD calculations for the anionic states, we 
added a set of 2s5p2d diffuse shells to the cc-pVTZ basis on 
all C and N atoms to achieve satisfactory stabilization of the 
$\eta$-trajectories of the resonance states. For those diffuse 
shells, the exponent ratios are 1.5 for the $p$ functions and 
2.0 for the $s$ and $d$ functions. The resulting basis set is 
identical to the one from Ref. \citenum{Ehara2017}. We used a 
cuboid CAP with onsets $r^0_x = 25.460$ a.u., $r^0_y = 7.039$ 
a.u., $ r^0_z = 5.047$ a.u. in all calculations. 

%%%%%%%%%%%%%%%%%%%%%%%%%%%%
% Results
%%%%%%%%%%%%%%%%%%%%%%%%%%%%%%%%%%%%%%%%%%%%%%%%%%%%%%%%%%%%%%%%%%%%%%%%
\section{Results}\label{sec:results}
\subsection{Energies and decay widths} \label{sec:res1}

The vertical electron affinity of fumaronitrile was determined 
as 0.93 eV using EOM-EA-CCSD/cc-pVTZ+2s5p2d, as compared to the 
SAC-CI value of 1.01 eV\cite{Ehara2017} and the experimental value 
for the adiabatic electron affinity of 1.21 eV.\cite{Khuseynov2014}  

\begin{figure*}[bt]
\captionsetup{justification=raggedright}
\includegraphics[width=0.45\textwidth]{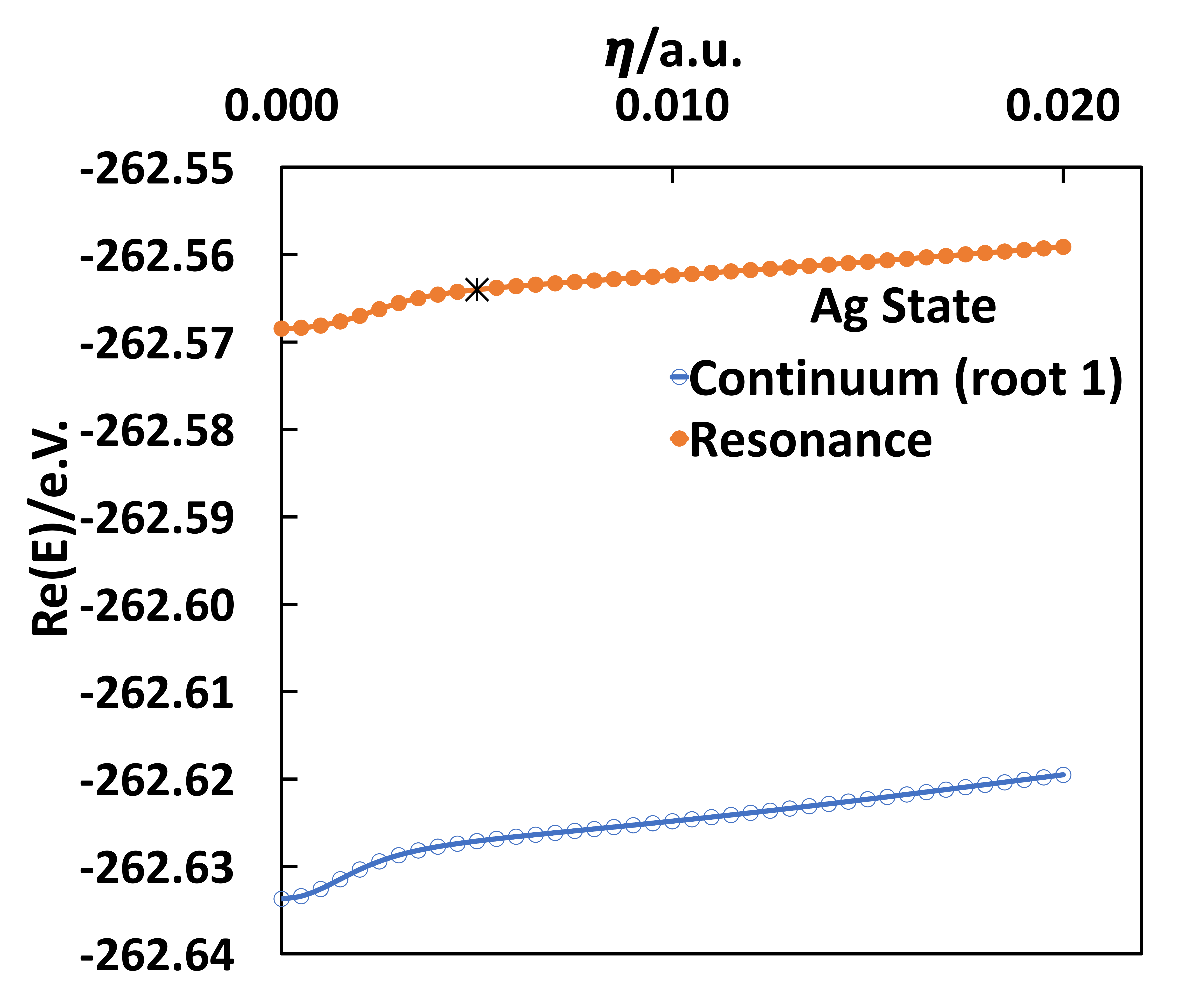} \hspace{0.5cm}
\includegraphics[width=0.45\textwidth]{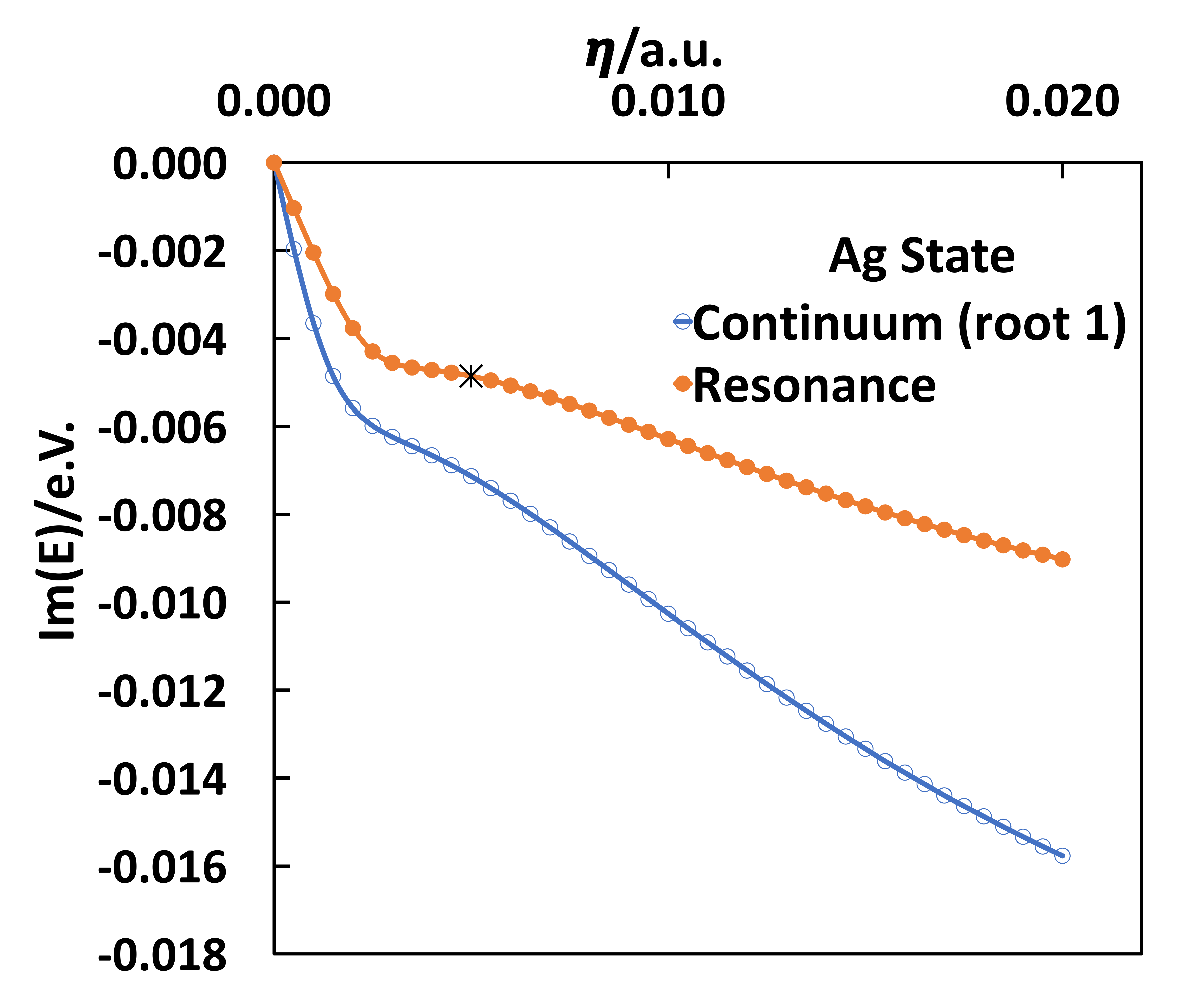} \\
\includegraphics[width=0.45\textwidth]{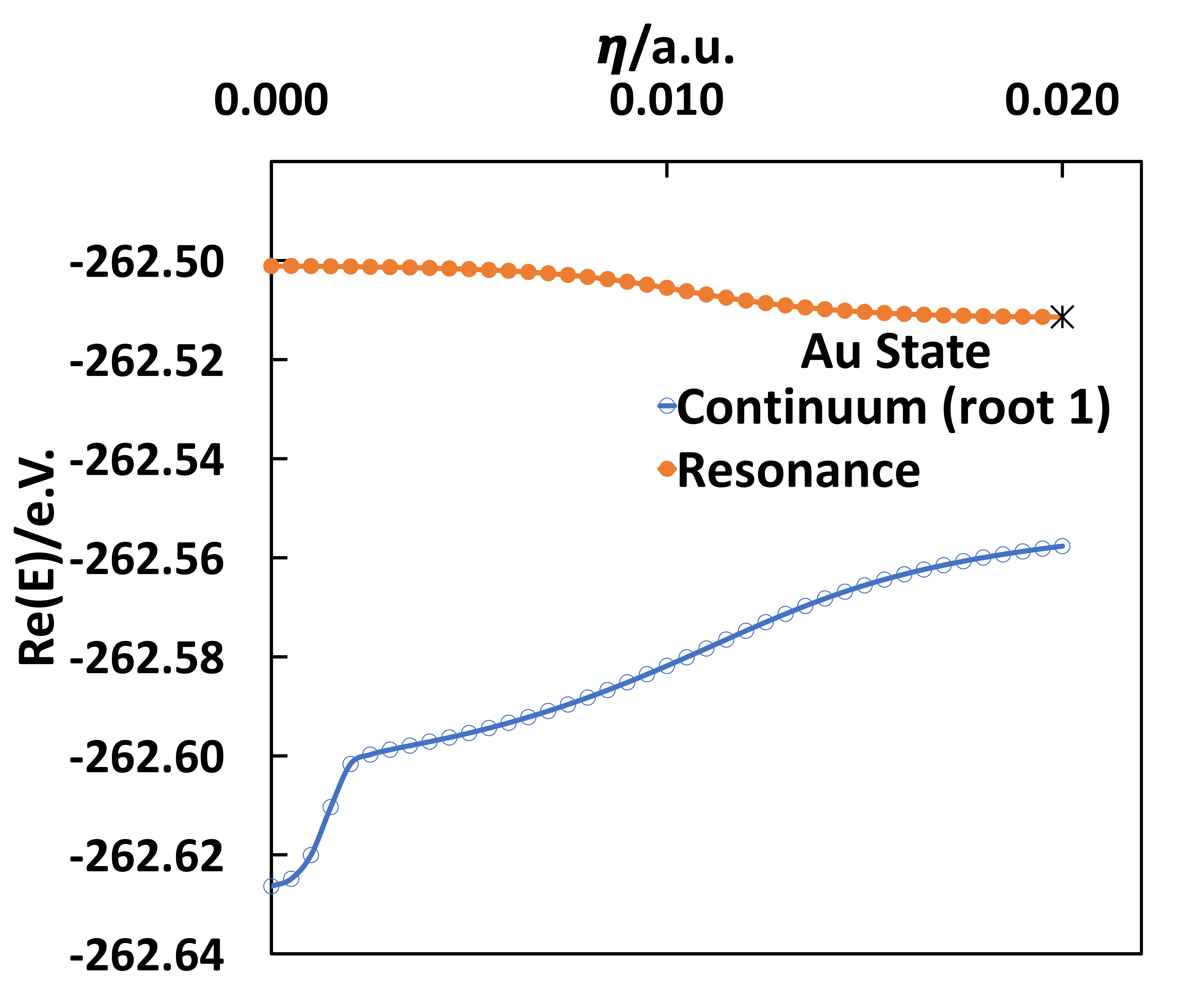} \hspace{0.5cm}
\includegraphics[width=0.45\textwidth]{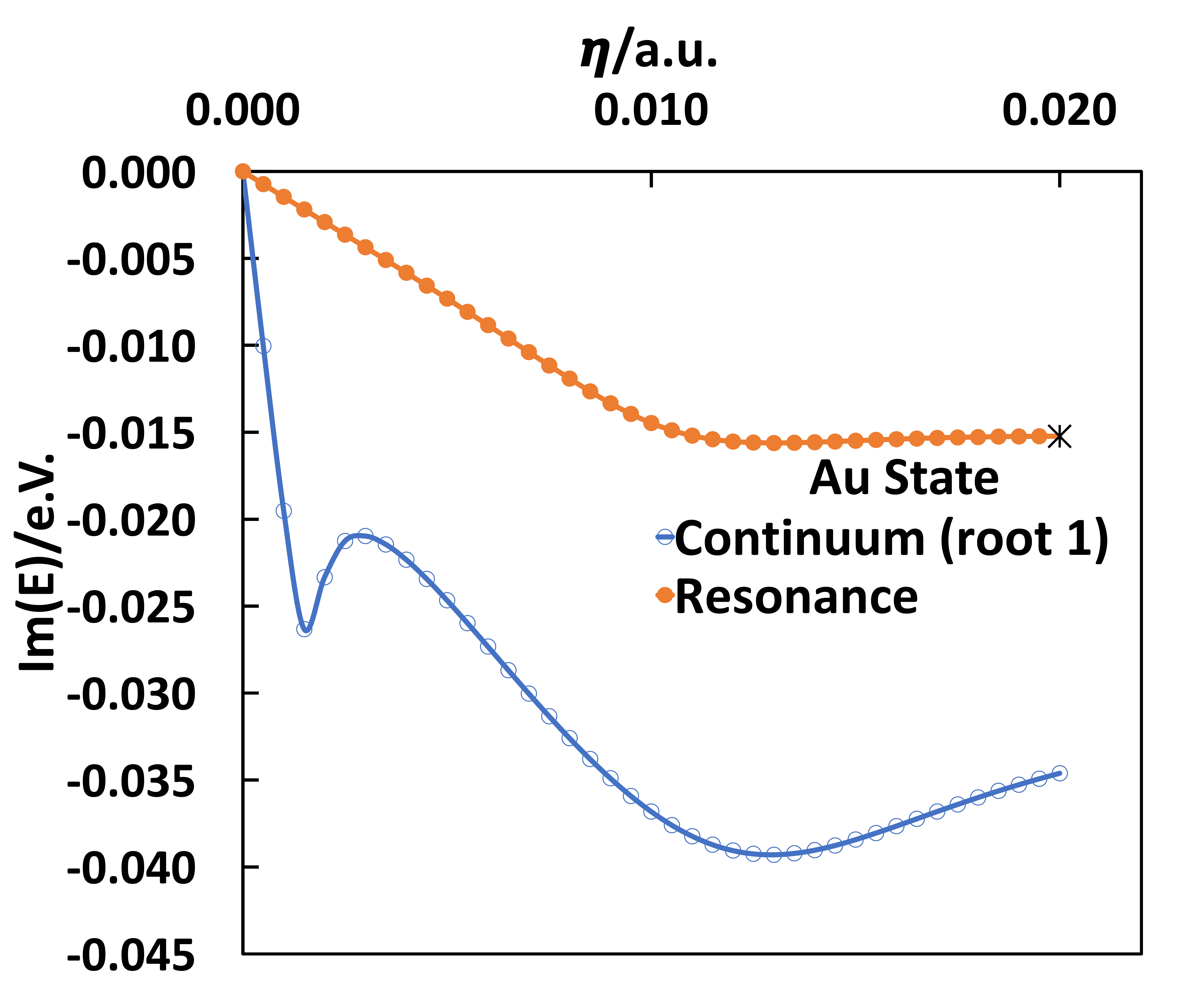}
\caption{Real (left) and imaginary (right) energies 
of resonance and pseudocontinuum states of $^2$A$_g$ (upper panels) 
and $^2$A$_u$ (lower panels) symmetry as a function of CAP strength 
$\eta$ computed with CAP-EOM-EA-CCSD/cc-pVTZ+2s5p2d at the equilibrium 
structure of neutral fumaronitrile. Optimal CAP 
strengths $\eta_\text{opt}$ are indicated by black stars.}
\label{fig:ag_au_res}
\end{figure*}

To identify the resonance states, we analyzed the behavior of the 
ten lowest CAP-EOM-EA-CCSD states of A$_g$ and A$_u$ symmetry as a 
function CAP strength $\eta$. \cref{fig:ag_au_res} shows the real 
and imaginary parts of the energy of the lowest-lying resonance 
and pseudocontinuum states of both symmetries. In both cases, the 
resonance state is energetically above the lowest pseudocontinuum 
state. \cref{fig:ag_au_res} also illustrates that the stabilization 
of the A$_u$ resonance is somewhat better than that of the A$_g$ 
resonance. From these $\eta$-trajectories, we determined the 
$\eta_{opt}$ values for the $^2$A$_g$ and $^2$A$_u$ resonances as 
0.005 a.u. and 0.020 a.u., respectively, by means 
of the criterion $\text{min}|\eta\, dE/d\eta|$ (see Sec. \ref{sec:th2}). 

In addition, we computed the positions and widths of these two 
resonance states with projected CAP-EOM-EA-CCSD.\cite{Gayvert2022} 
In these calculations, the CAP was constructed in a basis of 10 
EOM-EA-CCSD states; the optimal CAP strengths are 0.006 a.u. and 
0.021 a.u., respectively, for the $^2$A$_g$ and $^2$A$_u$ resonances. 
For both approaches, projected and full CAP-EOM-EA-CCSD, we also 
computed the first-order correction and analyzed the corresponding 
trajectories.\cite{Riss1993,Jagau2013} 

All computed resonance positions and widths are given in 
\cref{tab:resonance}. Full and projected CAP-EOM-EA-CCSD agree 
within 0.03 eV for the positions and widths of both states; only 
for the first-order corrected resonance position of the $^2$A$_u$ 
state, the difference is 0.1 eV. The  correction amounts to at 
most 0.06 eV. In view of this good agreement among different 
CAP-EOM-EA-CCSD variants, the substantial deviations from 
CAP-SAC-CI\cite{Ehara2017} that we observe for the $^2$A$_u$ 
resonance are somewhat surprising. For this state, the resonance 
position computed with CAP-EOM-EA-CCSD is 0.5 eV lower than the 
CAP-SAC-CI value, whereas the resonance width is about twice 
as large, i.e., 0.8 eV as compared to 0.4 eV. In contrast, we 
observe good agreement with CAP-SAC-CI for the $^2$A$_g$ 
resonance; the position and width differ by 
no more than 0.15 eV. 

\begin{figure*}[htb]
\subfloat[b$_g$ (bound)]{\includegraphics[width=0.25\textwidth]{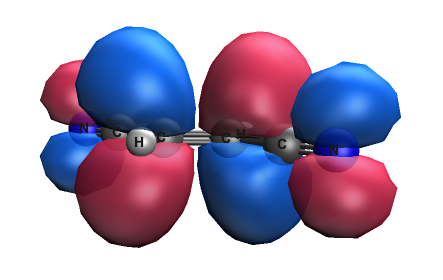}}
\captionsetup{justification=raggedright,singlelinecheck=false}
\subfloat[a$_g$ (resonance)]{\includegraphics[width=0.25\textwidth]{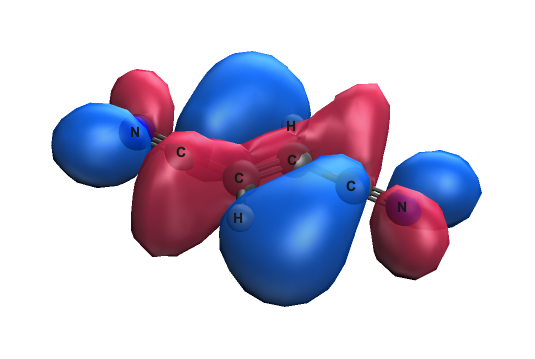}}
\captionsetup{justification=raggedright,singlelinecheck=false}
\subfloat[a$_u$ (resonance)]{\includegraphics[width=0.25\textwidth]{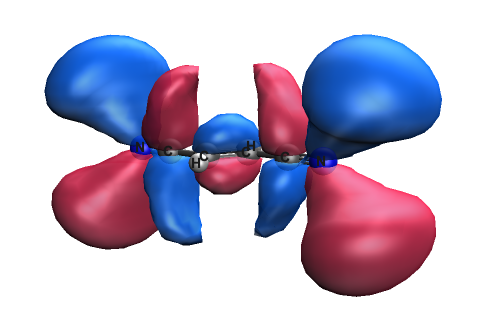}}
\captionsetup{justification=raggedright,singlelinecheck=false}
\caption{Real parts of Dyson orbitals for electron attachment to 
fumaronitrile computed with CAP-EOM-EA-CCSD/cc-pVTZ+2s5p2d at the 
respective $\eta_\text{opt}$ and plotted at an isovalue of 0.02.}
\label{fig:dyson_mo}
\end{figure*}

We also computed CAP-EOM-EA-CCSD Dyson orbitals\cite{Jagau2016} 
for the bound and temporary anion states of fumaronitrile. The 
real parts of these orbitals are shown in \cref{fig:dyson_mo}. 
It is evident that the b$_g$ and a$_u$ orbitals have out-of-plane 
character while the a$_g$ orbital has in-plane character. 

\begin{table*}[h] 
\caption{Resonance positions $E$ and widths $\Gamma$ of fumaronitrile 
in eV computed with full and projected CAP-EOM-EA-CCSD and projected 
CAP-SAC-CI. Values are given with and without the first-order correction.} 
\label{tab:resonance}
\begin{tabular}{lcccccc} \hline
State & \multicolumn{2}{c}{full CAP-EOM-EA-CCSD$^a$} & \multicolumn{2}{c}{
proj. CAP-EOM-EA-CCSD$^a$} & proj. CAP-SAC-CI$^b$ & Expt.$^c$ \\ \hline 
 & uncorr. & corr. & uncorr. & corr. &  &  \\ \hline
 & \multicolumn{6}{c}{Resonance positions} \\ \hline
$^2$A$_g$ & 2.27 & 2.21 & 2.28 & 2.24 & 2.35 & 1.8 \\
$^2$A$_u$ & 3.70 & 3.74 & 3.67 & 3.64 & 4.11 & 3.5 \\ \hline
 & \multicolumn{6}{c}{Resonance widths} \\ \hline 
$^2$A$_g$ & 0.26 & 0.31 & 0.27 & 0.33 & 0.39 & -- \\
$^2$A$_u$ & 0.80 & 0.82 & 0.78 & 0.84 & 0.37 & -- \\ \hline
\end{tabular}
{\footnotesize $^a$ This work. \\
$^b$ From Ref. \citenum{Ehara2017}, computed with an approximate 
symmetrized SAC-CI matrix and a smooth Voronoi CAP. \\
$^c$ From Ref. \citenum{Burrow1992}, determined using electron 
transmission spectroscopy.}
\end{table*}
%\footnotetext{This work.}
%\footnotetext{From Ref. \citenum{Ehara2017}, computed with an approximate 
%symmetrized SAC-CI matrix and a smooth Voronoi CAP.}
%\footnotetext{From Ref. \citenum{Burrow1992}, determined using electron 
%transmission spectroscopy.}

\subsection{Non-adiabatic coupling between bound and pseudocontinuum states} \label{sec:res2}

\begin{figure*}[htb]
%\subfloat[Coupling mode of Re${[h_z]}^{B_g-A_g}$.]{\label{fig:Bg_Ag_nac_Re500_continua}
%\includegraphics[width=0.45\textwidth]{Bg_Ag_nac_Re500_continua}}
%\captionsetup{justification=raggedright,singlelinecheck=false}
%\subfloat[Coupling mode of Im${[h_z]}^{B_g-A_g}$.]{\label{fig:Bg_Ag_nac_Im500_continua}
%\includegraphics[width=0.45\textwidth]{Bg_Ag_nac_Im500_continua}}
%\captionsetup{justification=raggedright,singlelinecheck=false}
\subfloat[Elements of $\text{Re}(h)$.]{\label{fig:bc1}
\includegraphics[width=0.45\textwidth]{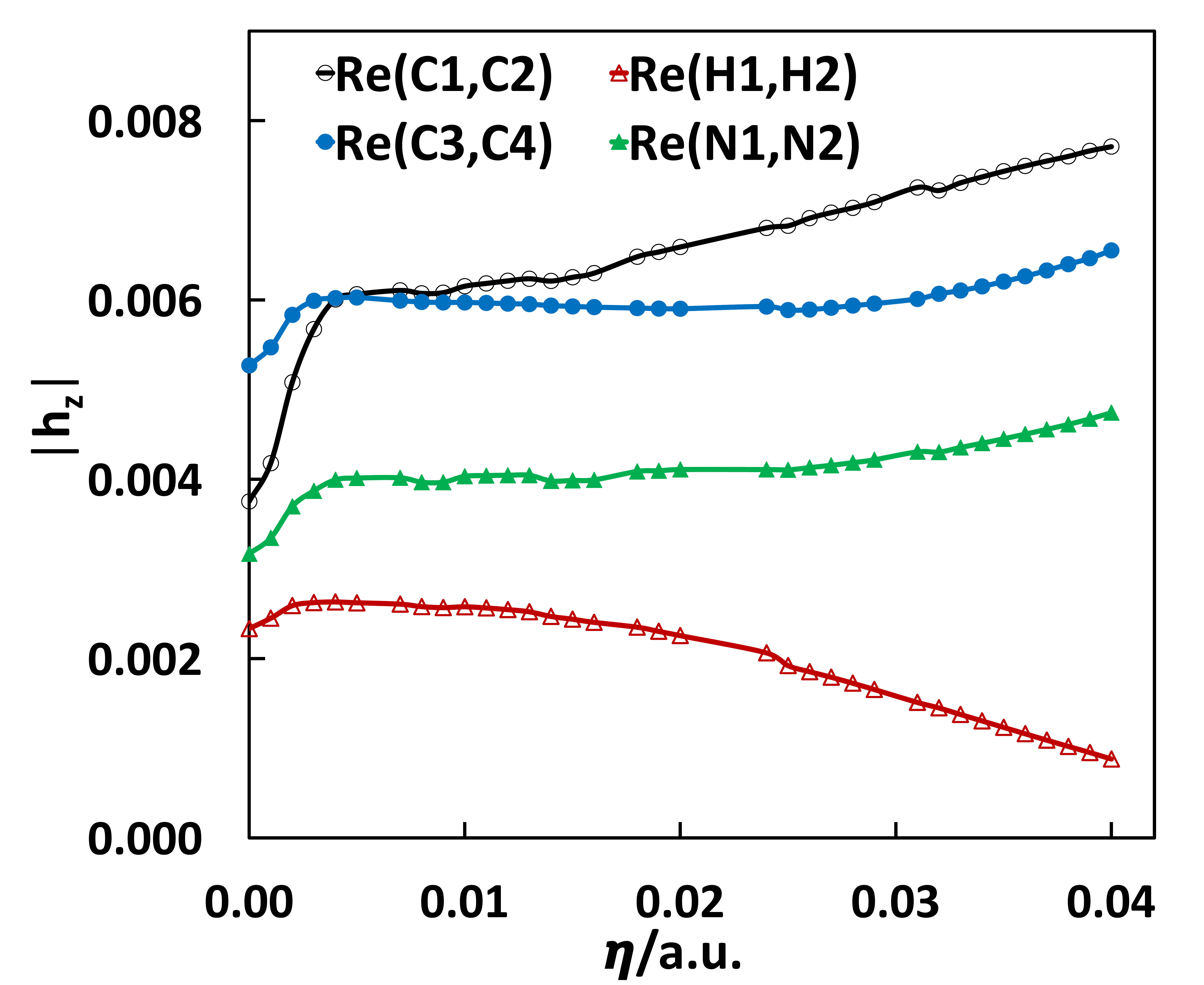}}
\captionsetup{justification=raggedright,singlelinecheck=false}
\subfloat[Elements of $\text{Im}(h)$.]{\label{fig:bc2}
\includegraphics[width=0.45\textwidth]{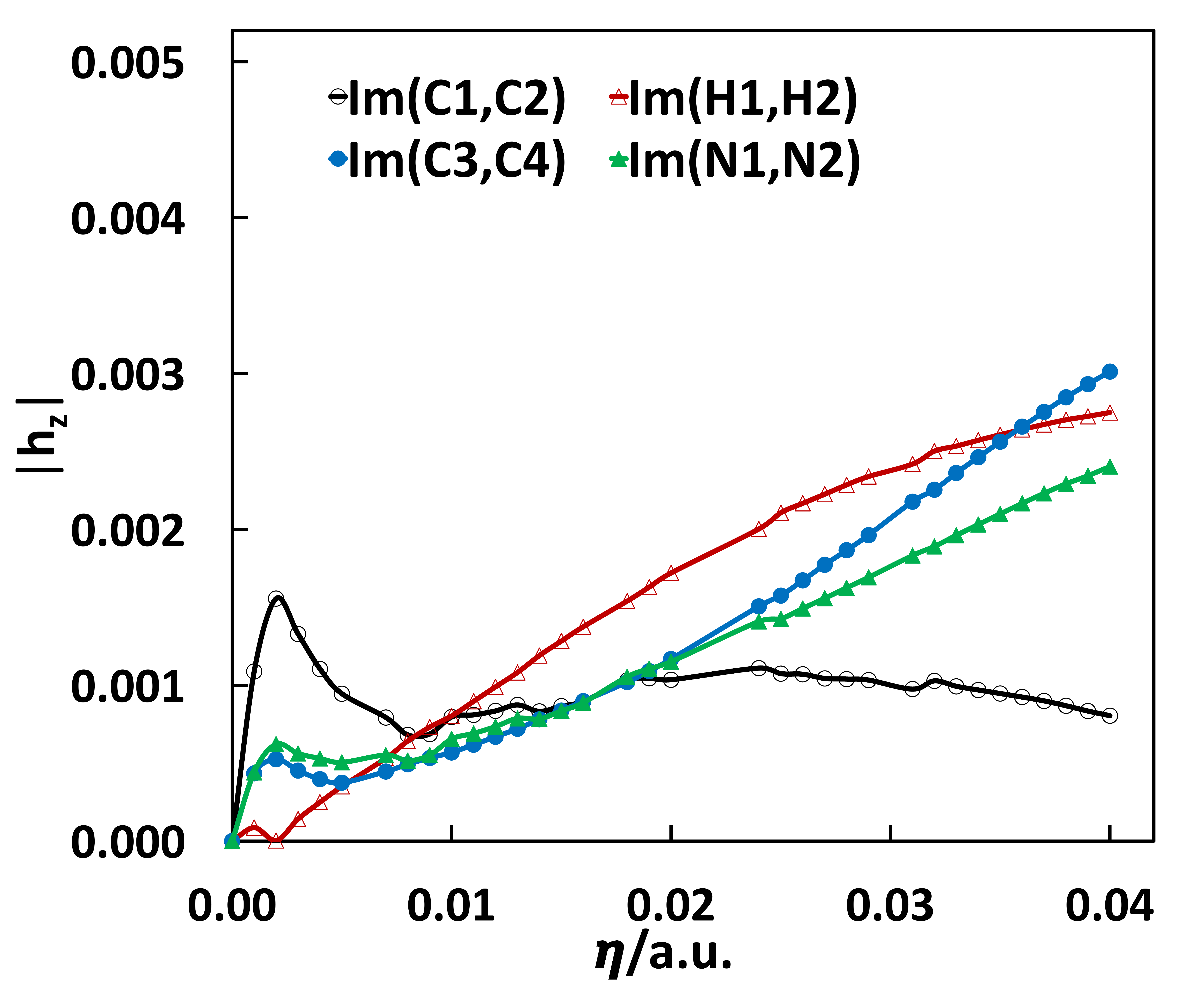}}
\captionsetup{justification=raggedright,singlelinecheck=false}
\subfloat[Norms of $\text{Re}(h)$ and $\text{Im}(h)$.]{\label{fig:bc3}
\includegraphics[width=0.45\textwidth]{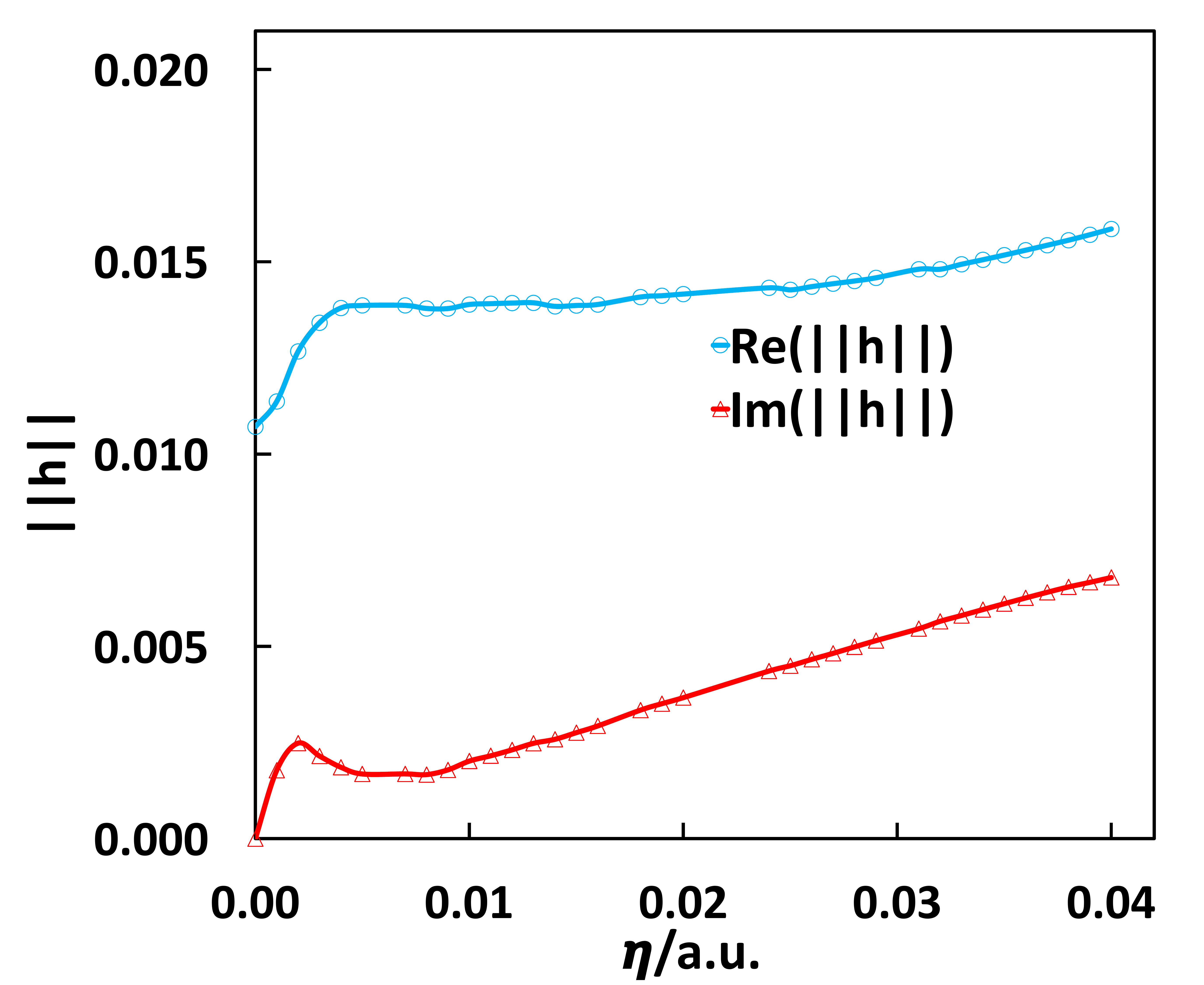}}
\captionsetup{justification=raggedright,singlelinecheck=false}
\subfloat[Norms of $\text{Re} (\mathcal{F})$ and $\text{Im} 
(\mathcal{F})$.]{\label{fig:bc4}
\includegraphics[width=0.45\textwidth]{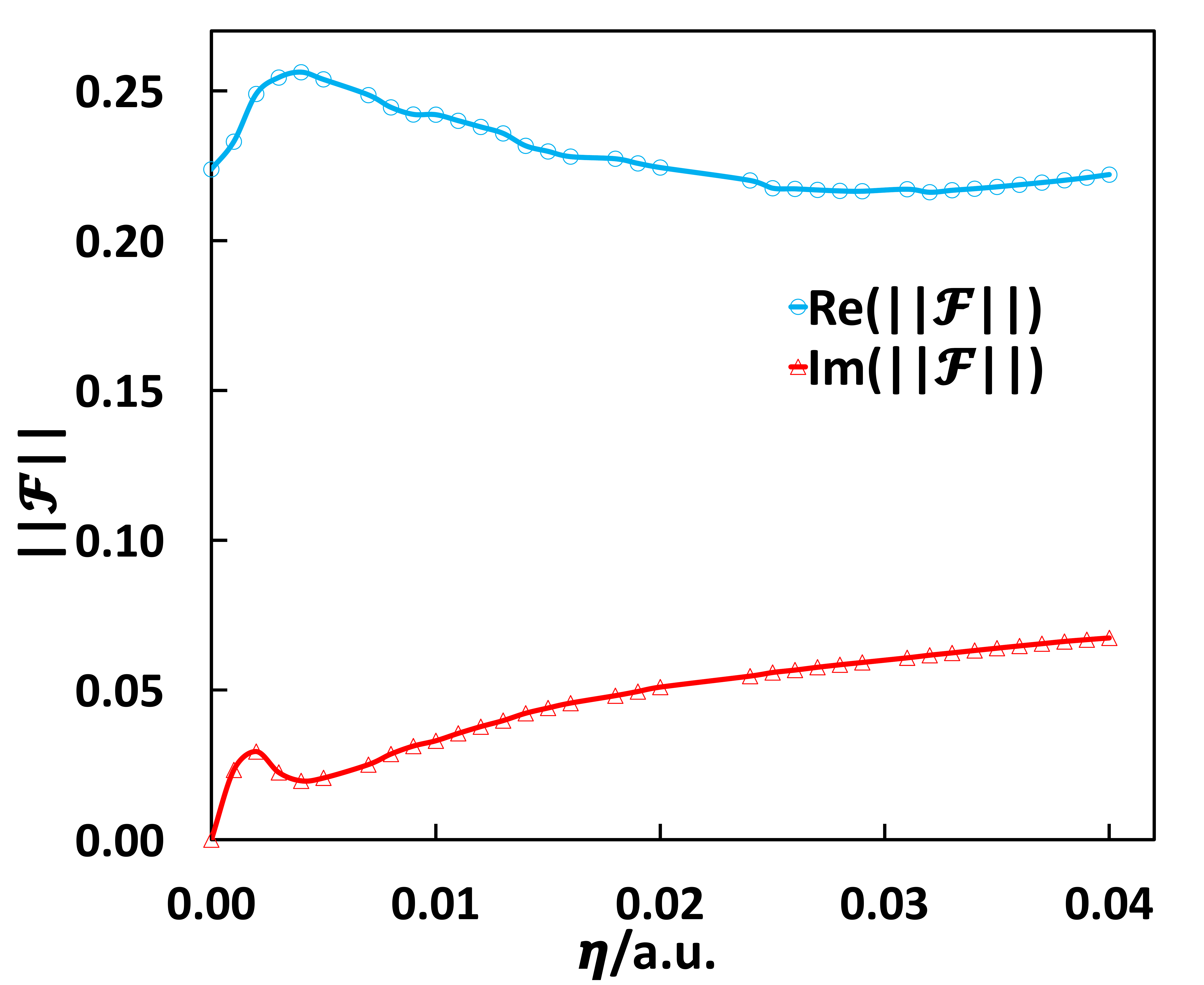}}
\captionsetup{justification=raggedright,singlelinecheck=false}
\caption{Non-adiabatic coupling force $h$ and derivative coupling 
$\mathcal{F}$ between the lowest pseudocontinuum state of A$_g$ 
symmetry and the bound $^2$B$_g$ state of fumaronitrile anion as 
a function of CAP strength $\eta$ computed at the equilibrium 
structure of the neutral molecule. See Sec. 
\ref{sec:compd} for explanation of the atom labels.}
\label{fig:bc}
\end{figure*}

Fig. \ref{fig:bc} shows the NAC force $h$ and the derivative coupling 
$\mathcal{F}$ between the lowest pseudocontinuum state of A$_g$ symmetry 
and the bound $^2$B$_g$ state of the fumaronitrile anion. Since the sign is 
arbitrary, we report the absolute values of all elements. We point out 
that the physical meaning of pseudocontinuum states in CAP theories is 
limited; the main purpose of Fig. \ref{fig:bc} is to enable a discussion 
of differences between pseudocontinuum states and resonances, which we 
do in the following sections. In addition, the analysis of coupling 
vectors involving pseudocontinuum states is helpful for the verification 
of our implementation. For example, the NAC force vector always needs to 
reflect spatial symmetry and the sum of all elements needs to vanish for 
couplings between bound, resonance, and pseudocontinuum states alike. 

The two states shown in Fig. \ref{fig:bc} are coupled by vibrations 
of $b_g$ symmetry, which correspond to out-of-plane motions in $z$ 
direction. Panels $a$ and $b$ of Fig. \ref{fig:bc} illustrate the 
dependence of the four symmetry-unique elements of $h$ on the CAP 
strength $\eta$, while panels $c$ and $d$ show the same for the norms 
of $h$ and $\mathcal{F}$, respectively. It is apparent that there is 
no stabilization with respect to $\eta$ in the imaginary part of either 
$h$ or $\mathcal{F}$, which reflects the behavior of the energy of 
pseudocontinuum states. The asymptotically linear dependence on $\eta$ 
according to Eq. \eqref{eq:cxnac} is clearly visible. Somewhat 
surprisingly, the real part of $h$ does not vary much in the range 
$\eta = 0.005-0.020$ a.u.; only at larger CAP strengths the dependence 
is more pronounced. Notably, the four elements of $h$ are all of the 
same order of magnitude, while their dependence on $\eta$ differs a 
little. 

%%%%%%%%%%%%%%%%%%%%%%%%%%%%%%%%%%%%%%%%%%%%%%%%%%%%%%%%%%%%%%%%%%%%%%%%%

\subsection{Non-adiabatic coupling between bound and resonance states}
\label{sec:res3}

\begin{figure*}[h!]
\subfloat[Real part of NAC force at $\eta_\text{opt}$ = 0.005.]
{\label{fig:brA1}
\includegraphics[width=0.45\textwidth]{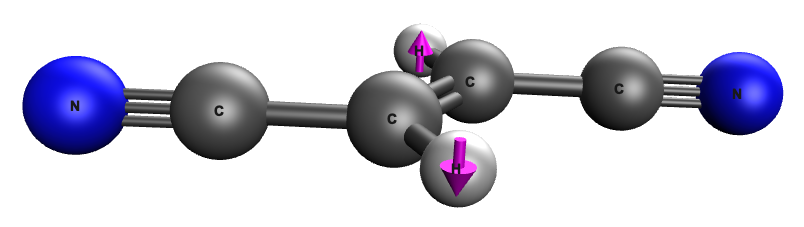}}
\captionsetup{justification=raggedright,singlelinecheck=false}
\subfloat[Imaginary part of NAC force at $\eta_\text{opt}$ = 0.005.]
{\label{fig:brA2}
\includegraphics[width=0.45\textwidth]{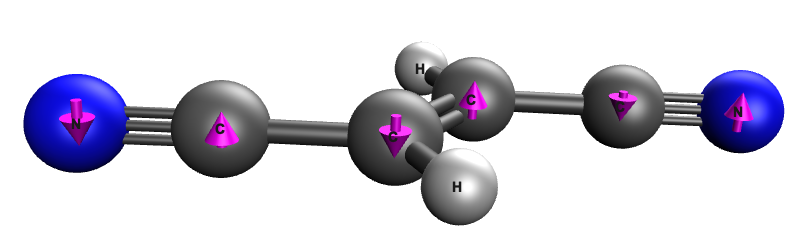}}
\captionsetup{justification=raggedright,singlelinecheck=false}
\subfloat[Elements of $\text{Re}(h)$.]{\label{fig:brA3}
\includegraphics[width=0.45\textwidth]{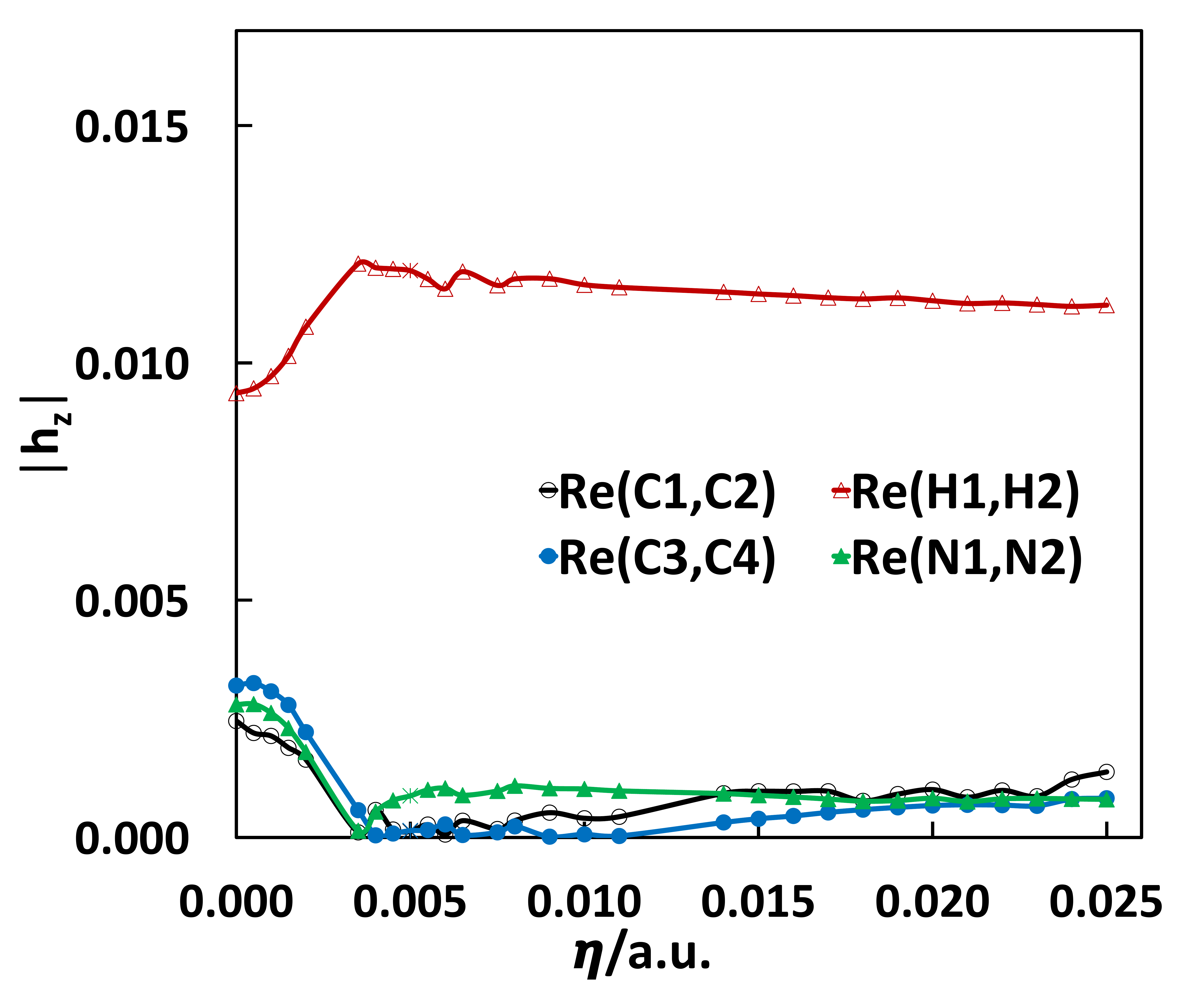}}
\captionsetup{justification=raggedright,singlelinecheck=false}
\subfloat[Elements of $\text{Im}(h)$.]{\label{fig:brA4}
\includegraphics[width=0.45\textwidth]{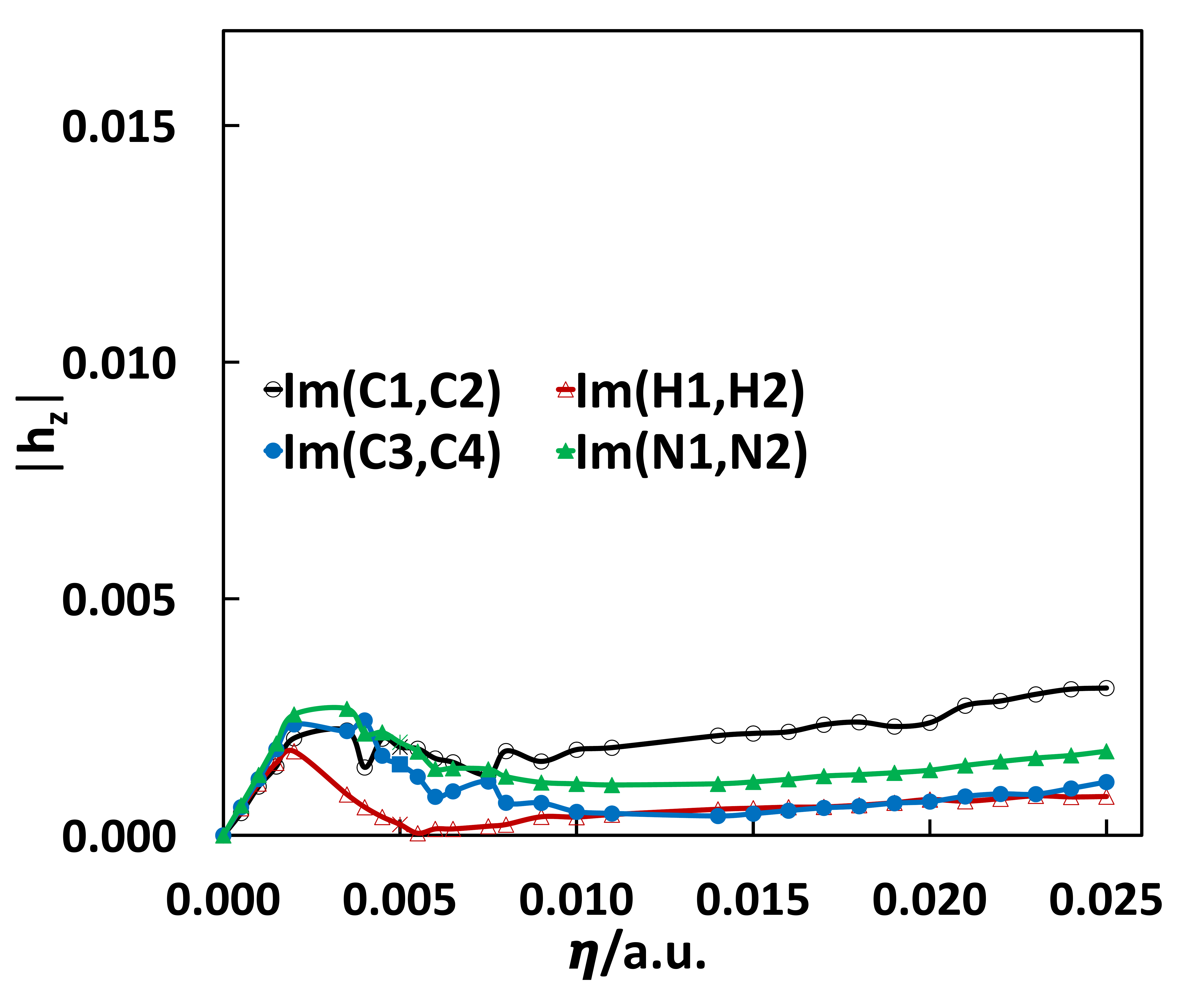}}
\captionsetup{justification=raggedright,singlelinecheck=false}
\subfloat[Norms of $\text{Re}(h)$ and $\text{Im}(h)$.]{\label{fig:brA5}
\includegraphics[width=0.45\textwidth]{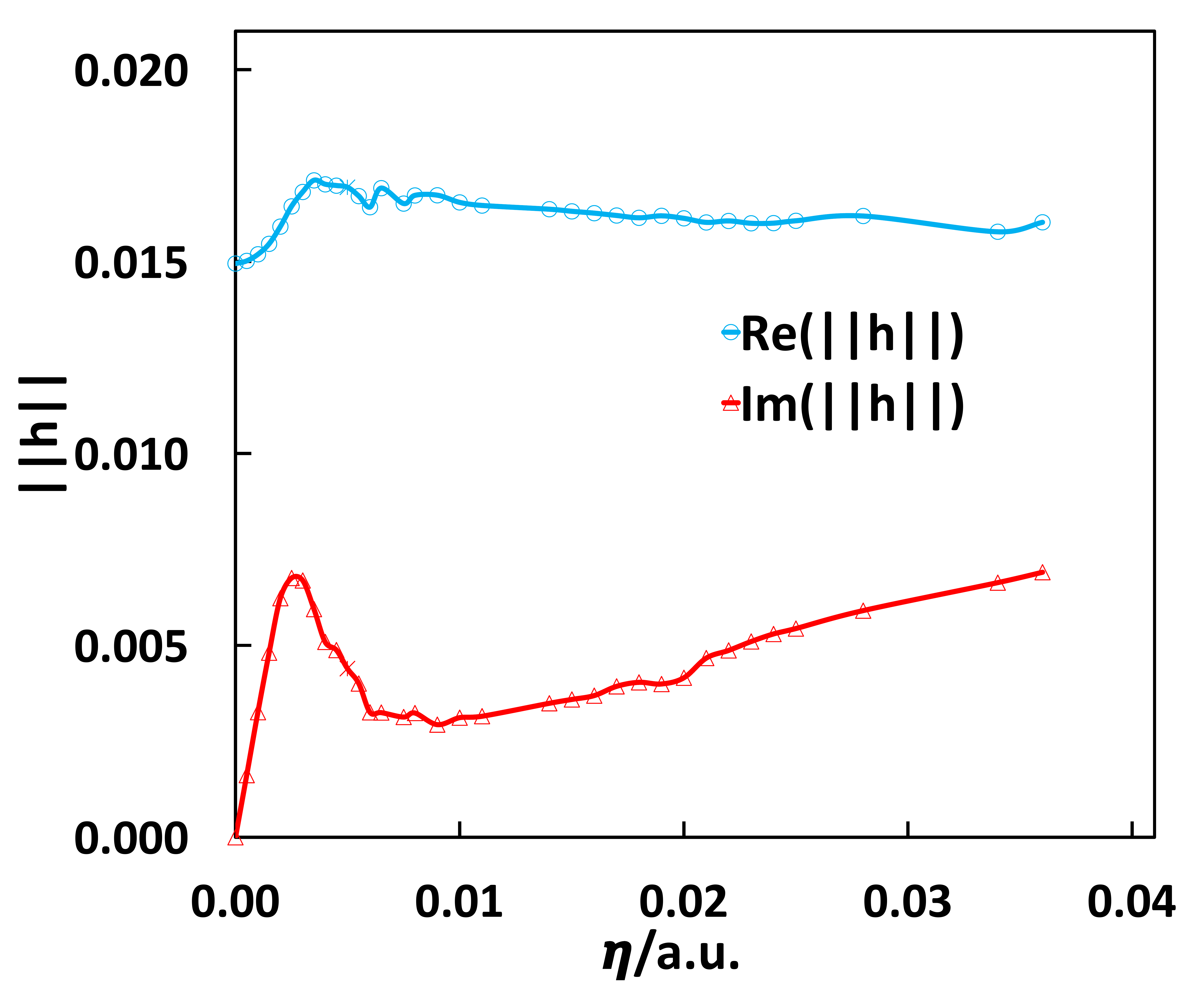}}
\captionsetup{justification=raggedright,singlelinecheck=false}
\subfloat[Norms of $\text{Re}(\mathcal{F})$ and $\text{Im}
(\mathcal{F})$.]{\label{fig:brA6}
\includegraphics[width=0.45\textwidth]{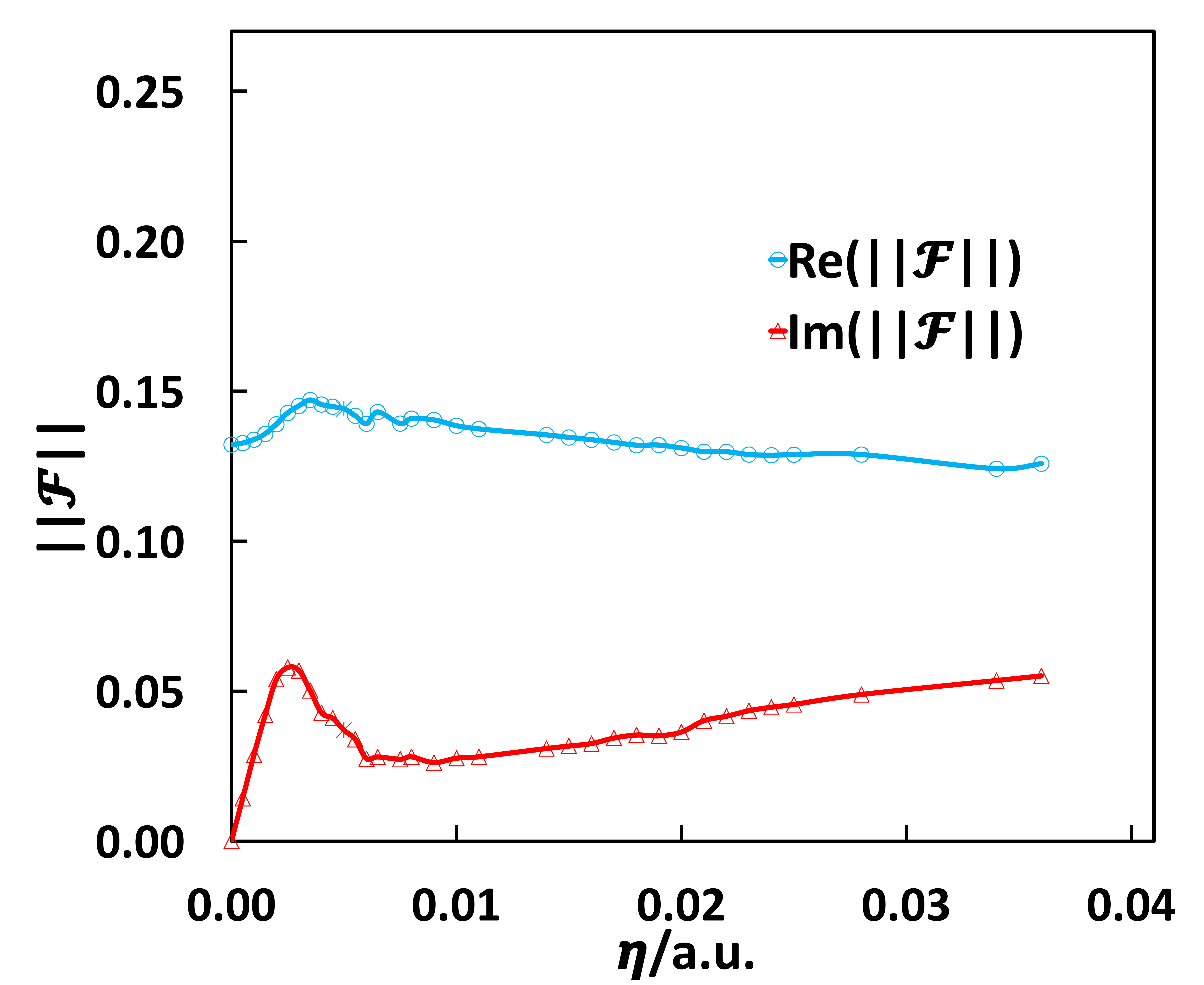}}
\captionsetup{justification=raggedright,singlelinecheck=false}
\caption{Non-adiabatic coupling force $h$ and derivative coupling 
$\mathcal{F}$ between the $^2$A$_g$ resonance and the bound $^2$B$_g$ 
state of fumaronitrile anion as a function of CAP strength $\eta$ 
computed at the equilibrium structure of the neutral molecule. 
See Sec. \ref{sec:compd} for explanation of the 
atom labels.}
\label{fig:brA}
\end{figure*} 

The NAC force and the derivative coupling between the bound $^2$B$_g$ 
state and the $^2$A$_g$ resonance are shown in Fig. \ref{fig:brA}. 
As the irreducible representations are the same as for the pair of 
states displayed in Fig. \ref{fig:bc}, the same four symmetry-unique 
elements of $h$ and $\mathcal{F}$ are non-zero. However, whereas all 
four elements are of similar magnitude for the coupling to the 
pseudocontinuum state in Fig. \ref{fig:bc}, the coupling to the 
resonance state is dominated by one single element, the out-of-plane 
motion of the hydrogen atoms. This element of $\text{Re}(h)$ is about 
10 times larger than the other three elements of $\text{Re}(h)$ and 
all four elements of $\text{Im}(h)$. Although the linear dependence 
on $\eta$ is visible for large $\eta$ values, all elements of $h$ 
vary less with $\eta$ above $\eta_\text{opt}=0.005$ a.u. than in 
Fig. \ref{fig:bc}, especially as concerns the imaginary part. 
This reflects the behavior of the energy and 
illustrates that the resonance is stabilized.

It is thus meaningful to evaluate the NAC force at one particular CAP 
strength as done in panels $a$ and $b$ of Fig. \ref{fig:brA}. This 
graphics illustrates that $\text{Re}(h)$ and $\text{Im}(h)$ point in 
different directions. More specifically, the angle between these two 
vectors is $97^\circ$. $\text{Re}(h)$ and $\text{Re}(\mathcal{F})$ 
are nearly collinear, whereas $\text{Im}(h)$ and $\text{Im}(\mathcal{F})$ 
span an angle of ca. $9^\circ$. Although it is difficult 
to assign physical meaning to these angles, we repeat that $h$ and 
$\mathcal{F}$ are necessarily collinear in Hermitian quantum chemistry, 
while this is not the case in non-Hermitian quantum chemistry by virtue 
of Eq. \eqref{eq:nacvec}.

The norms of $h$ and $\mathcal{F}$ are displayed in panels $e$ and $f$ 
of Fig. \ref{fig:brA}. Notably, the differences between the pseudocontinuum 
state and the resonance are less visible here than in the individual 
elements of $h$. It is also worth noting that the evaluation of $h$ 
and $\mathcal{F}$ at $\eta=0$, i.e., with bound-state EOM-CCSD, gives 
different results: Although the norms are similar, the ratio between 
the elements is markedly different at $\eta_\text{opt}$.

\begin{figure*}[h!] 
\subfloat[Real part of NAC force at $\eta_\text{opt}$ = 0.02.]
{\label{fig:brB1}
\includegraphics[width=0.45\textwidth]{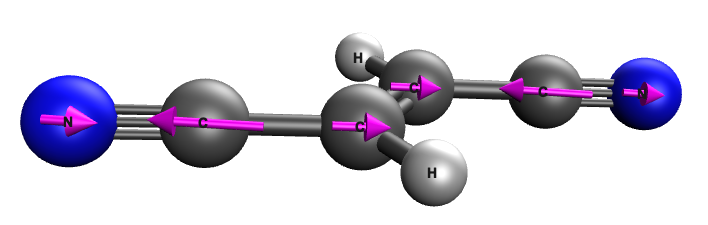}}
\captionsetup{justification=raggedright,singlelinecheck=false}
\subfloat[Imaginary part of NAC force at $\eta_\text{opt}$ = 0.02.]
{\label{fig:brB2}
\includegraphics[width=0.45\textwidth]{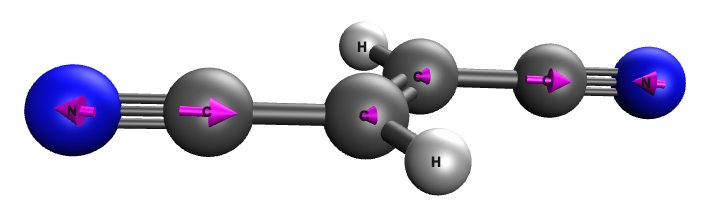}}
\captionsetup{justification=raggedright,singlelinecheck=false}
\subfloat[Elements of $\text{Re}(h)$, motion in $x$ direction.]
{\label{fig:brB3}
\includegraphics[width=0.45\textwidth]{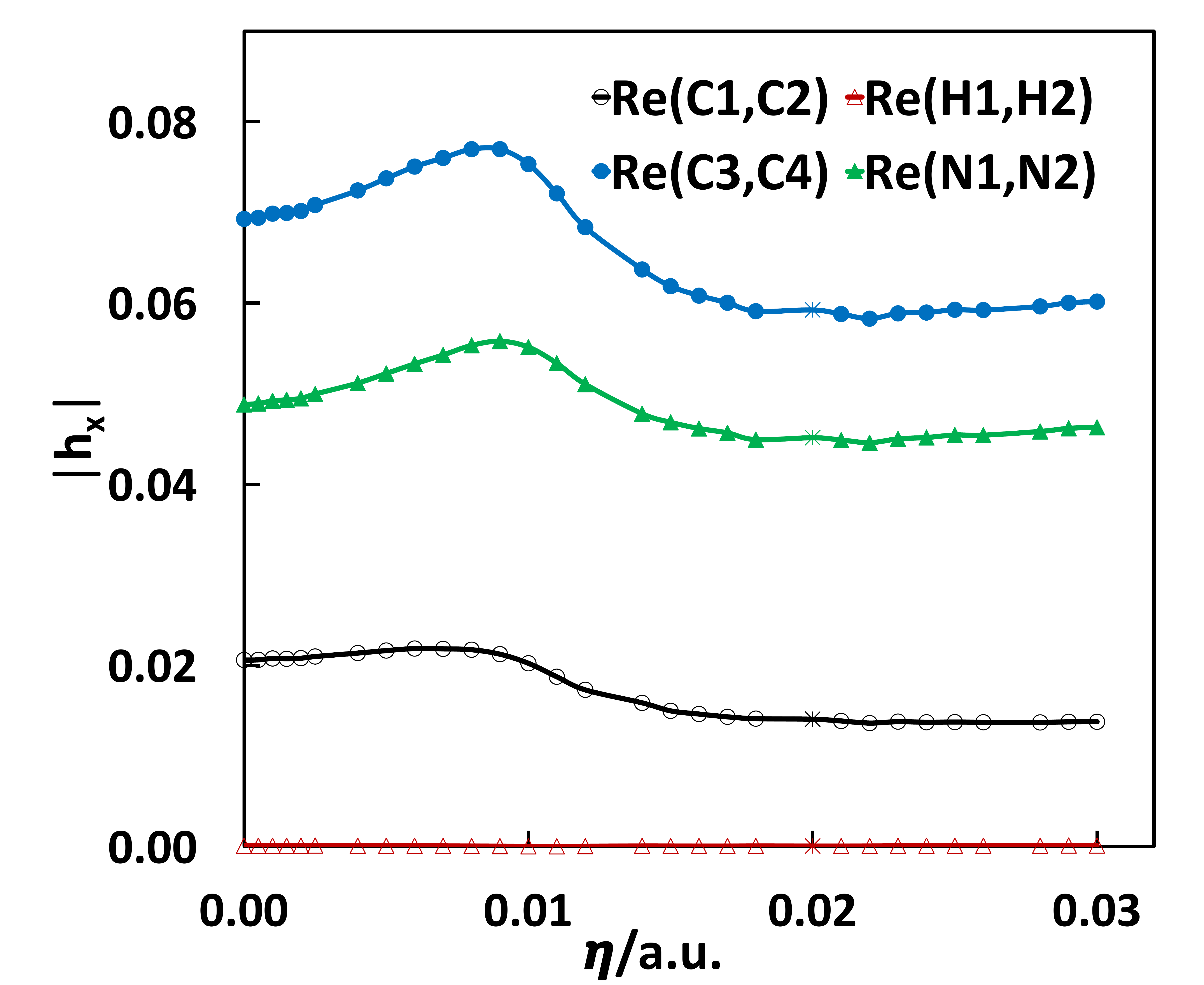}}
\captionsetup{justification=raggedright,singlelinecheck=false}
\subfloat[Elements of $\text{Im}(h)$, motion in $x$ direction.]
{\label{fig:brB4}
\includegraphics[width=0.45\textwidth]{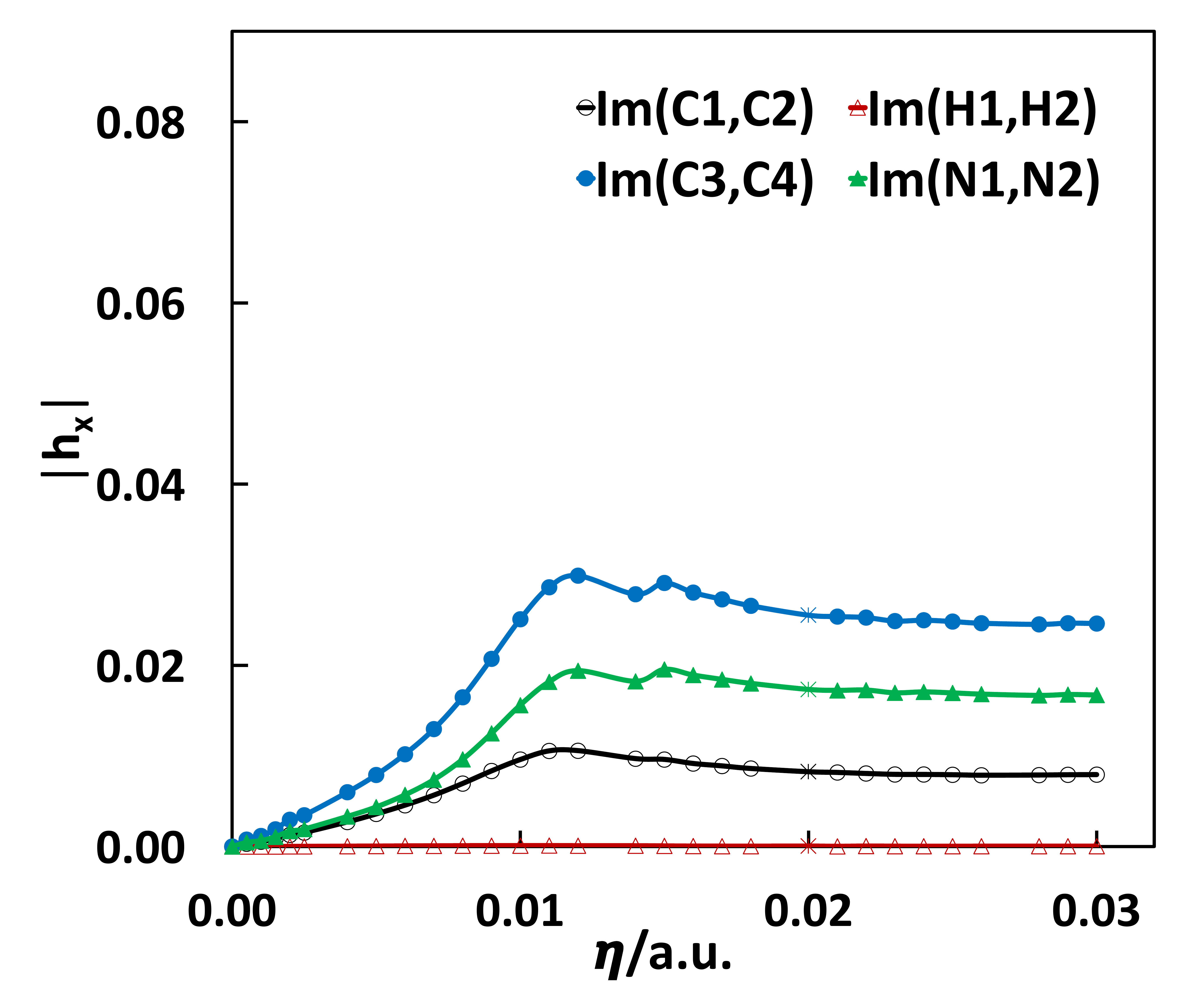}}
\captionsetup{justification=raggedright,singlelinecheck=false}
\subfloat[Elements of $\text{Re}(h)$, motion in $y$ direction.]
{\label{fig:brB5}
\includegraphics[width=0.45\textwidth]{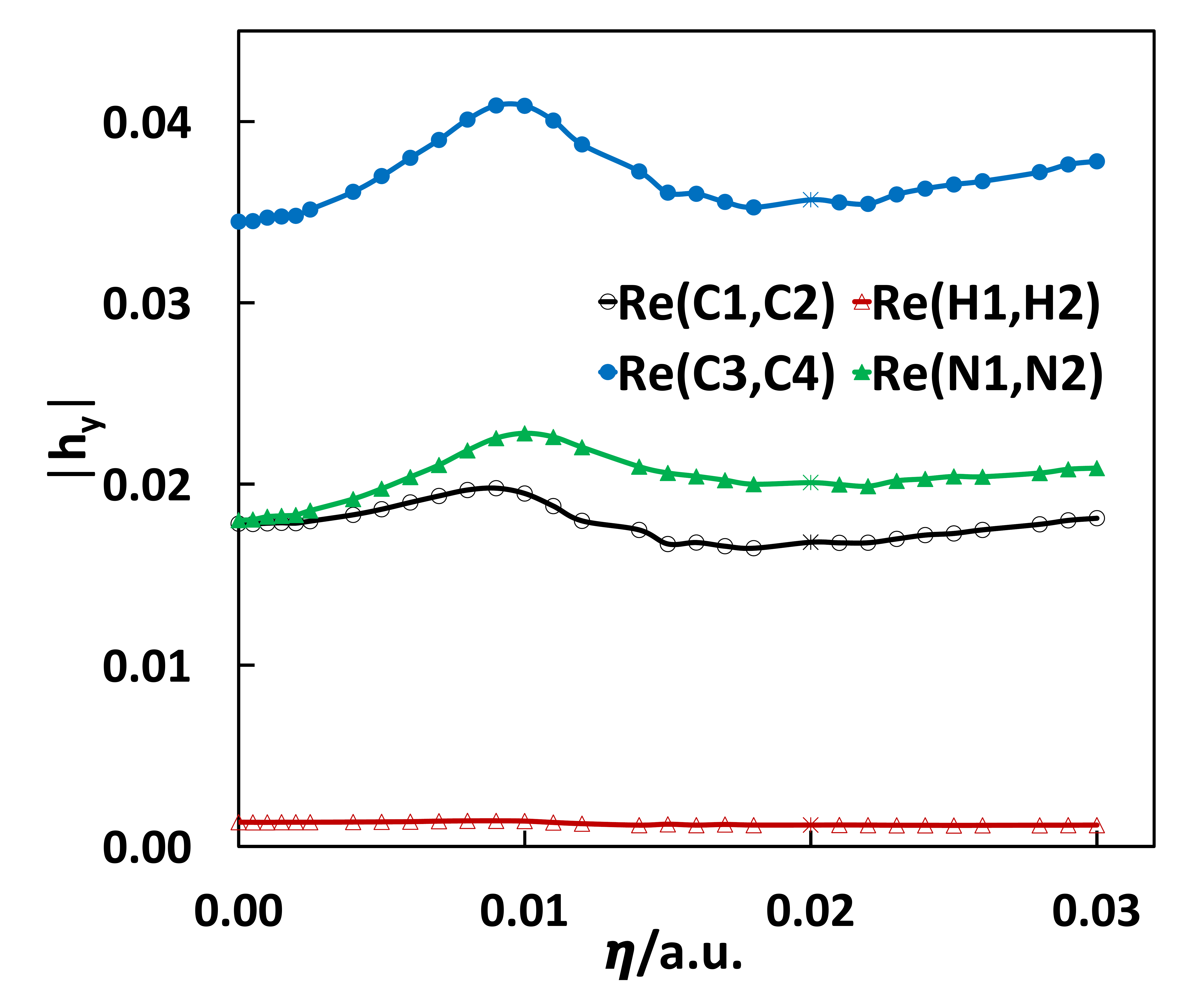}}
\captionsetup{justification=raggedright,singlelinecheck=false}
\subfloat[Elements of $\text{Im}(h)$, motion in $y$ direction.]
{\label{fig:brB6}
\includegraphics[width=0.45\textwidth]{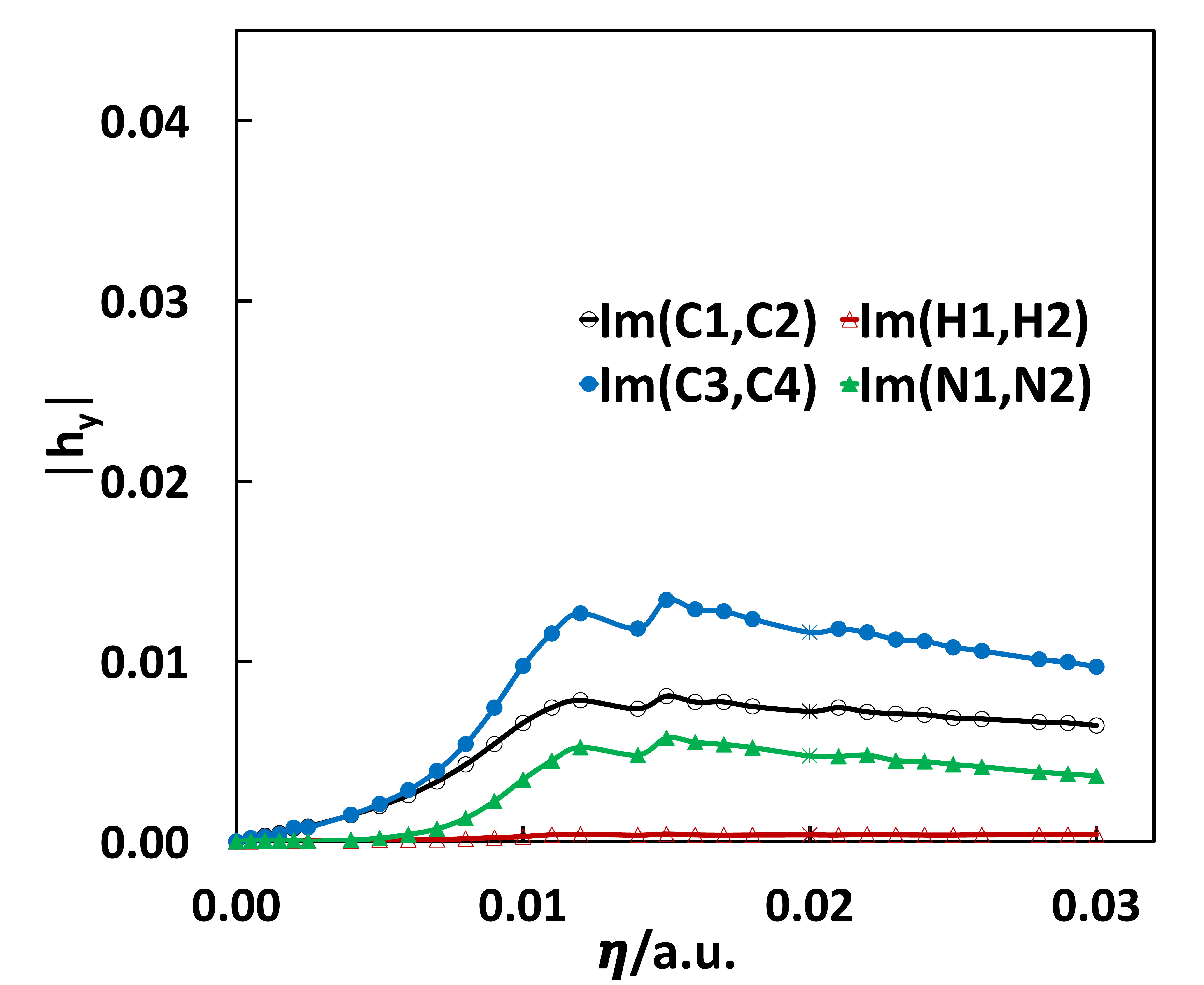}}
\captionsetup{justification=raggedright,singlelinecheck=false}
\caption{Non-adiabatic coupling force $h$ between the $^2$A$_u$ 
resonance and the bound $^2$B$_g$ state of fumaronitrile anion 
as a function of CAP strength $\eta$ computed at the equilibrium 
structure of the neutral molecule. See Sec. 
\ref{sec:compd} for explanation of the atom labels.}
\label{fig:brB}
\end{figure*}

\begin{figure*}[h!]
\subfloat[Norm of $\text{Re}(h)$.]{\label{fig:brC1}
\includegraphics[width=0.45\textwidth]{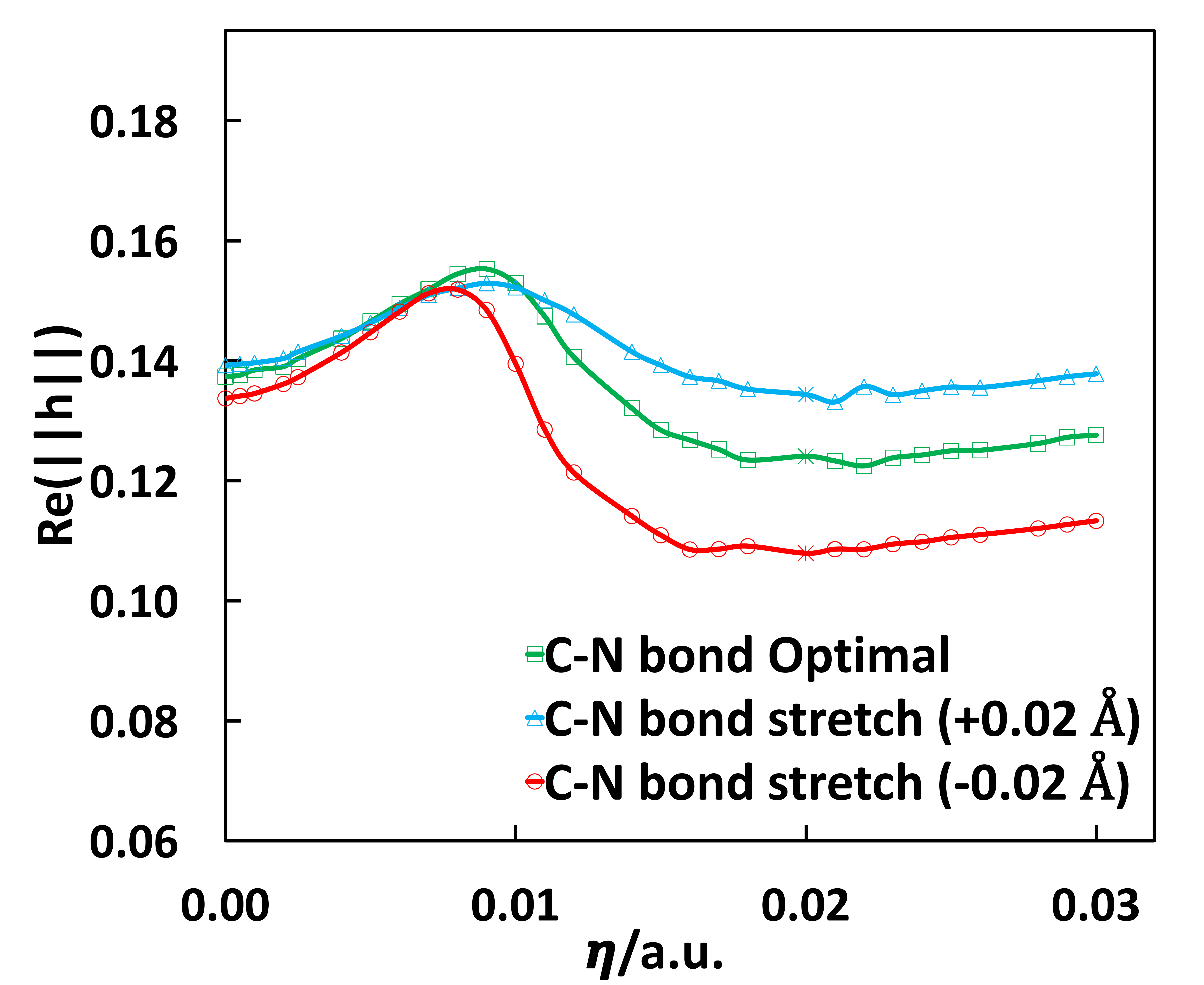}}
\subfloat[Norm of $\text{Im}(h)$.]{\label{fig:brC2}
\includegraphics[width=0.45\textwidth]{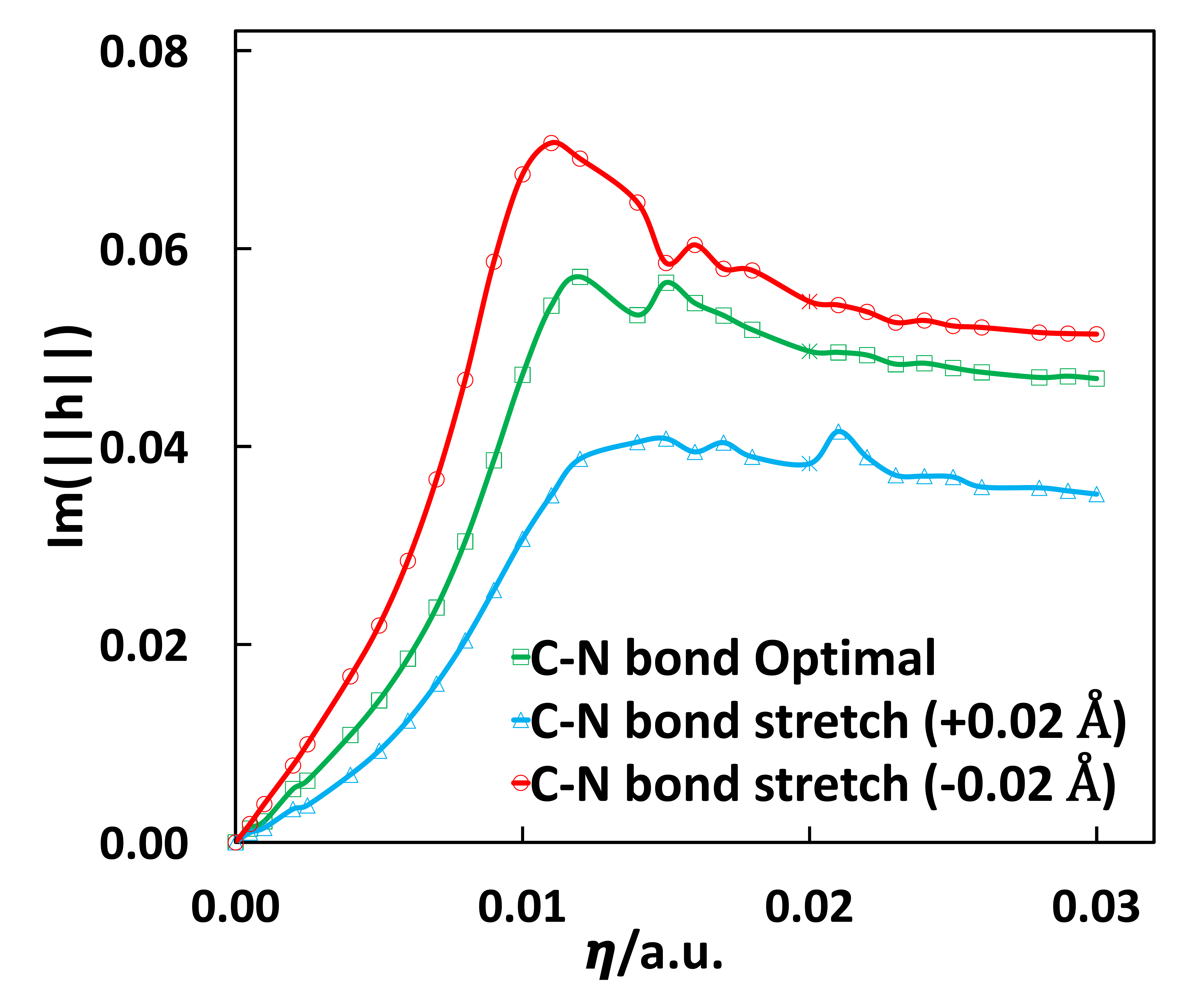}}
\captionsetup{justification=raggedright,singlelinecheck=false}
\subfloat[Norm of $\text{Re}(\mathcal{F})$.]{\label{fig:brC3}
\includegraphics[width=0.45\textwidth]{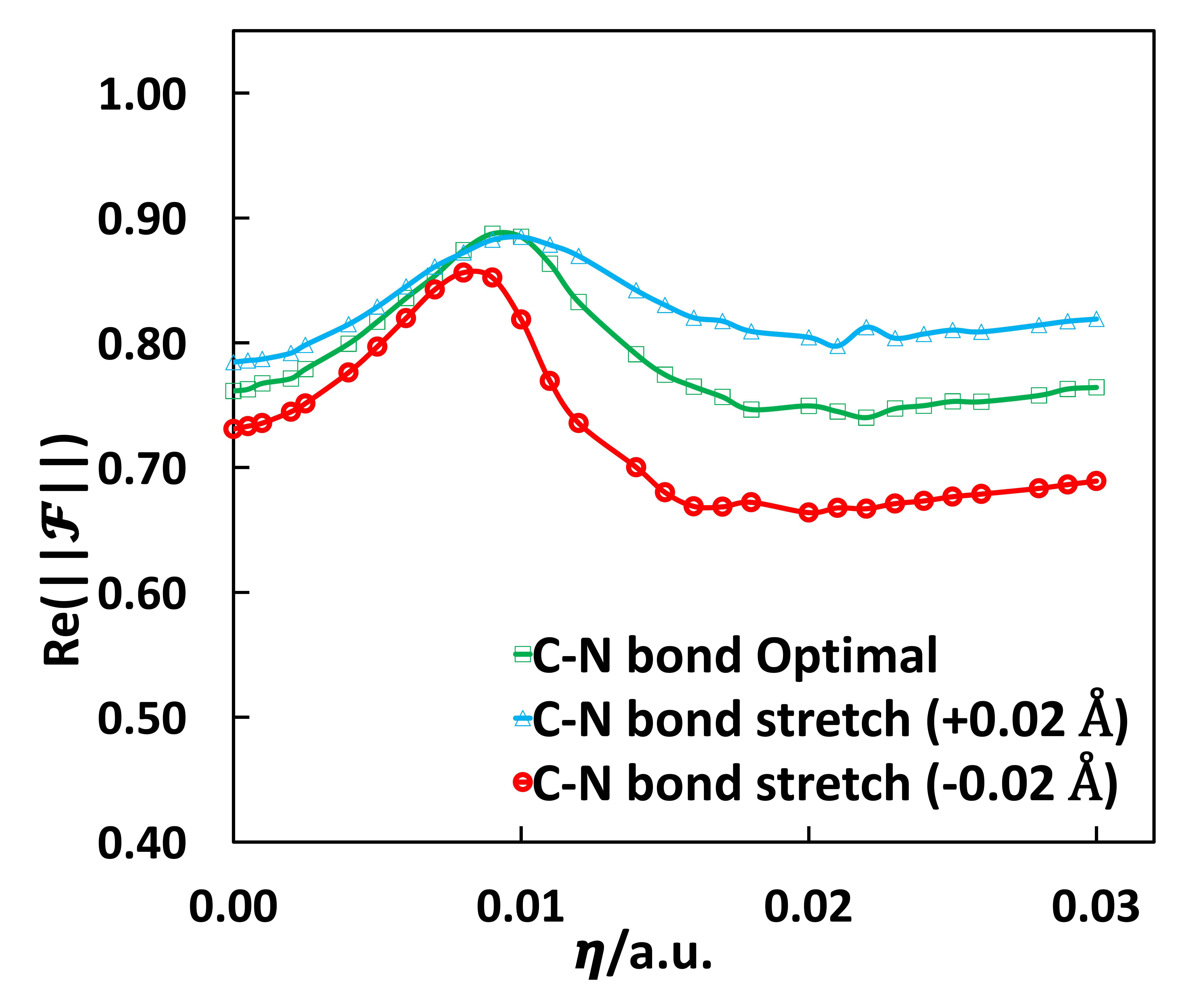}}
\captionsetup{justification=raggedright,singlelinecheck=false}
\subfloat[Norm of $\text{Im}(\mathcal{F})$.]{\label{fig:brC4}
\includegraphics[width=0.45\textwidth]{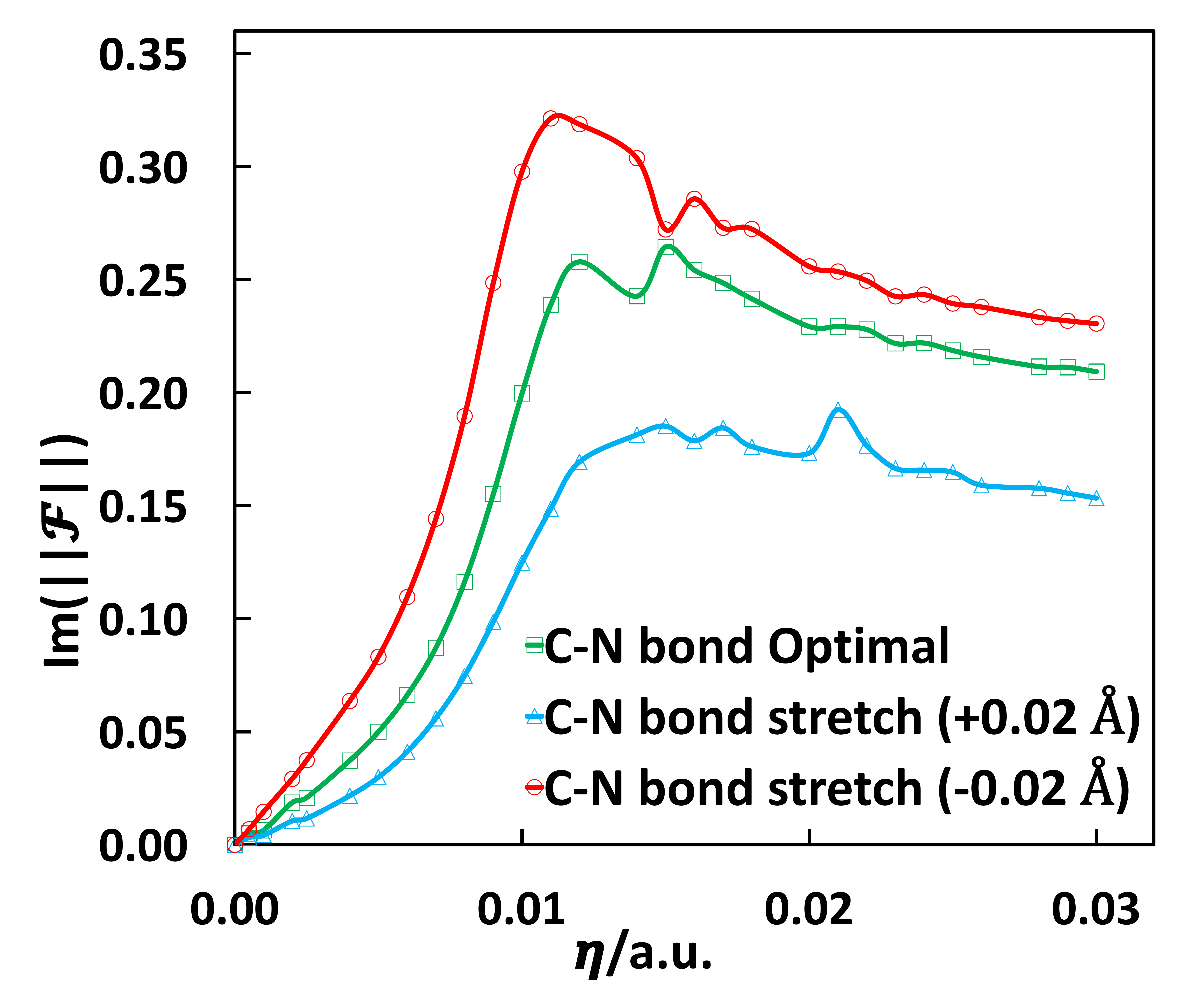}}
\captionsetup{justification=raggedright,singlelinecheck=false}
\caption{Non-adiabatic coupling force $h$ and derivative coupling 
$\mathcal{F}$ between the $^2$A$_u$ resonance and the bound 
$^2$B$_g$ state of fumaronitrile anion as a function of CAP 
strength $\eta$ at different structures.}
\label{fig:brC}
\end{figure*}

As a second example, shown in Figs. \ref{fig:brB} and \ref{fig:brC}, 
we investigated the NAC between the bound $^2$B$_g$ state and the 
$^2$A$_u$ resonance. These two states are coupled by vibrations of 
$b_u$ symmetry, which correspond to motions in the $xy$ plane. There 
are eight symmetry-unique elements of $h$, whose dependence on $\eta$ 
is displayed in Fig. \ref{fig:brB}. Similar to Fig. \ref{fig:brA}, 
all elements of $\text{Re}(h)$ and $\text{Im}(h)$ do not change much 
with $\eta$ above the optimal value, which is 0.020 a.u. for the 
$^2$A$_u$ resonance. The two elements of $h$ that correspond to the 
motion of the hydrogen atoms are almost zero, whereas the atoms of 
the cyano groups have the largest elements in $h$. Similar to the 
previous example, $\text{Re}(h)$ and $\text{Im}(h)$ point in different 
directions but here the angle between them amounts to ca. $171^\circ$. 
$\text{Re}(h)$ and $\text{Re}(\mathcal{F})$ are again almost collinear 
and the angle between $\text{Im}(h)$ and $\text{Im}(\mathcal{F})$ is 
ca. $2^\circ$.

An important difference between the two resonance states is that the 
excitation from the bound $^2$B$_g$ state to the $^2$A$_u$ resonance 
is bright, whereas the excitation to the $^2$A$_g$ resonance is dark 
owing to spatial symmetry. We would thus expect that the NAC between 
the $^2$B$_g$ and $^2$A$_u$ states could be probed in a photodetachment 
experiment on the anion of fumaronitrile if direct and indirect 
detachment can be distinguished. 

For this reason, we investigated the dependence of $h$ and $\mathcal{F}$ 
on the CN bond distance for the coupling between the $^2$B$_g$ and 
$^2$A$_u$ states. Fig. \ref{fig:brC} demonstrates that $h$ and $\mathcal{F}$ 
are very sensitive to the molecular structure as already a small change 
of 0.02 \AA\ in the CN distance changes the norms of $\text{Re}(h)$ 
and $\text{Im}(h)$ by more than 10\%. Interestingly, the norm of 
$\text{Re}(h)$ grows when the CN bond is stretched while the norm 
of $\text{Im}(h)$ shrinks. We note that the width of the $^2$A$_u$ 
state also changes significantly from 0.96 eV over 0.80 eV to 0.63 eV 
when stretching the bond. However, the energy gap between the two states 
only changes from 4.60 eV to 4.63 eV as both anionic states are stabilized 
with respect to the neutral ground state when the CN bond is stretched. 
This can be explained by the shape of the Dyson orbitals shown in Fig. 
\ref{fig:dyson_mo}, both of which have nodal planes across the CN bond.

%%%%%%%%%%%%%%%%%%%%%%%%%%%%%%%%%%%%%%%%%%%%%%%%%%%%%%%%%%%%%%%%%%%%%%%%%

\subsection{Non-adiabatic coupling between two resonance states}
\label{sec:res4}

\begin{figure*}[h!]
\subfloat[Real part of NAC force at $\eta_\text{opt}$ = 0.005.]{\label{fig:rr1}
\includegraphics[width=0.45\textwidth]{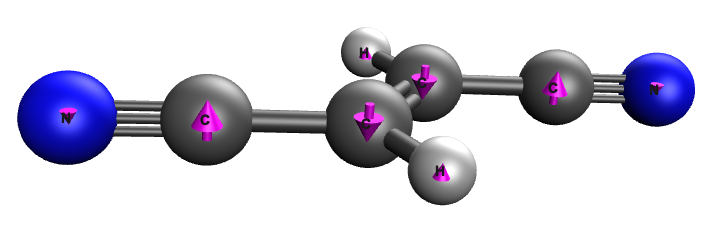}}
\captionsetup{justification=raggedright,singlelinecheck=false}
\subfloat[Imaginary part of NAC force at $\eta_\text{opt}$ = 0.005.]{\label{fig:rr2}
\includegraphics[width=0.45\textwidth]{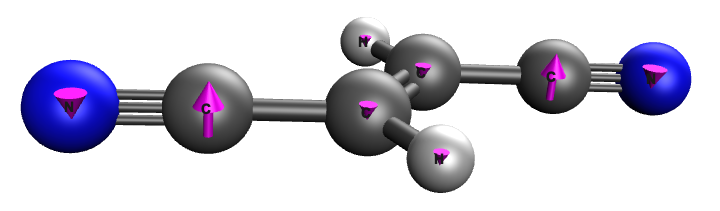}}
\captionsetup{justification=raggedright,singlelinecheck=false}
\subfloat[Real part of NAC force at $\eta_\text{opt}$ = 0.02.]{\label{fig:rr3}
\includegraphics[width=0.45\textwidth]{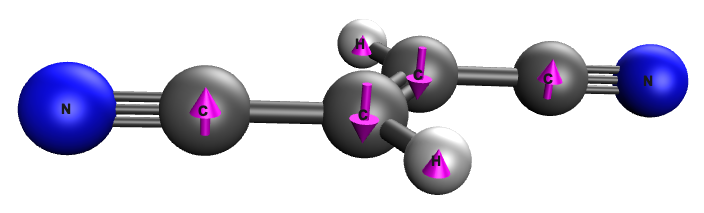}}
\captionsetup{justification=raggedright,singlelinecheck=false}
\subfloat[Imaginary part of NAC force at $\eta_\text{opt}$ = 0.02.]{\label{fig:rr4}
\includegraphics[width=0.45\textwidth]{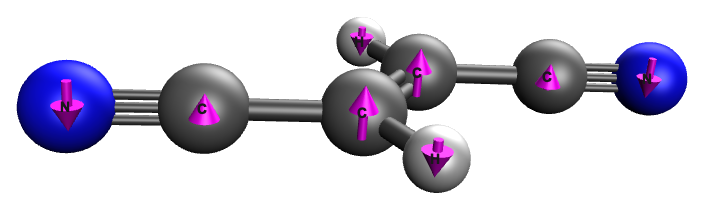}}
\captionsetup{justification=raggedright,singlelinecheck=false}
\subfloat[Elements of $\text{Re}(h)$.] {\label{fig:rr5}
\includegraphics[width=0.45\textwidth]{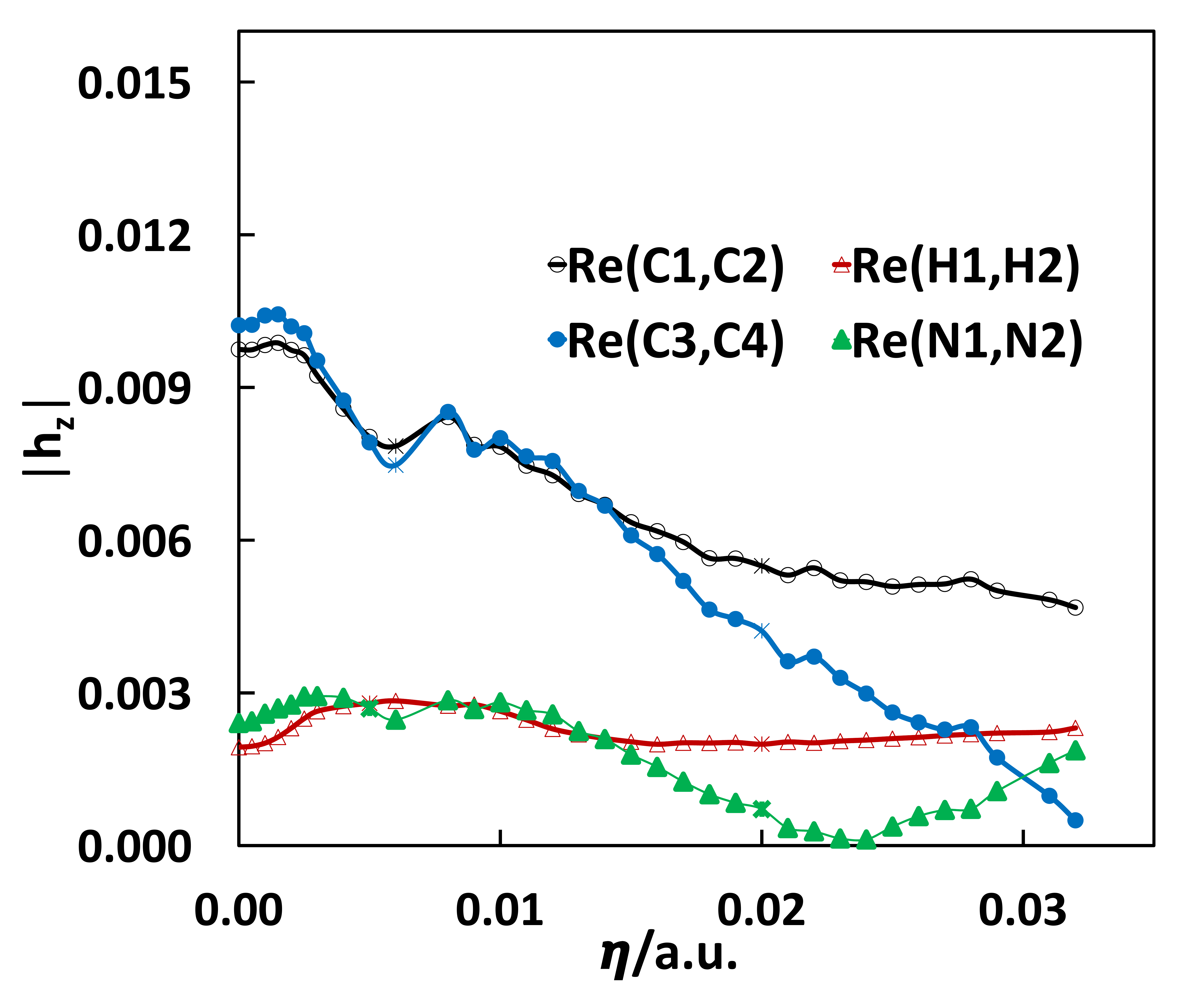}}
\captionsetup{justification=raggedright,singlelinecheck=false}
\subfloat[Elements of $\text{Im}(h)$.] {\label{fig:rr6}
\includegraphics[width=0.45\textwidth]{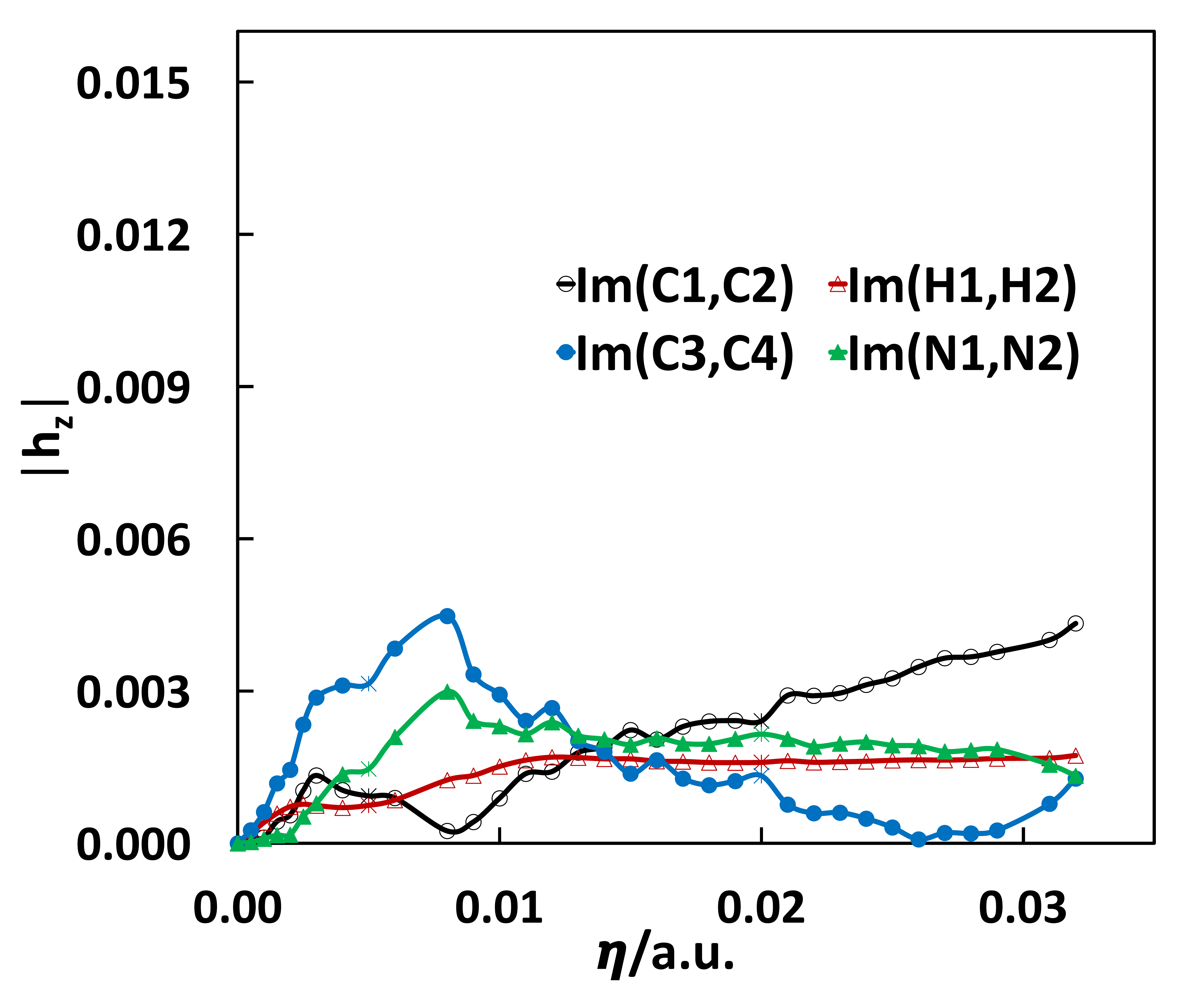}}
\captionsetup{justification=raggedright,singlelinecheck=false}
\subfloat[Norms of $\text{Re}(h)$ and $\text{Im}(h)$.]{\label{fig:rr7}
\includegraphics[width=0.45\textwidth]{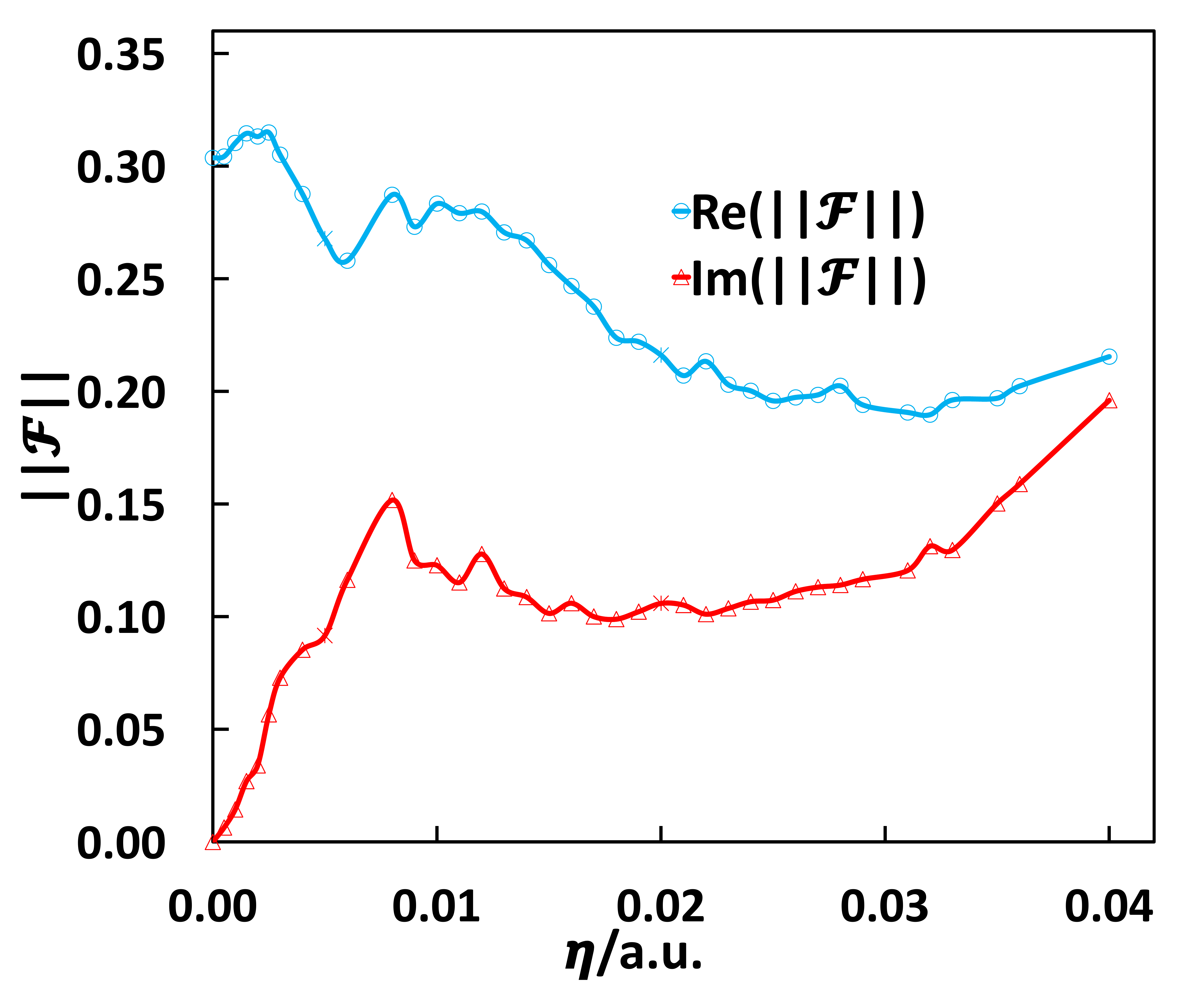}}
\captionsetup{justification=raggedright,singlelinecheck=false}
\subfloat[Norms of $\text{Re}(\mathcal{F}$ and $\text{Im}(\mathcal{F})$.]
{\label{fig:rr8}
\includegraphics[width=0.45\textwidth]{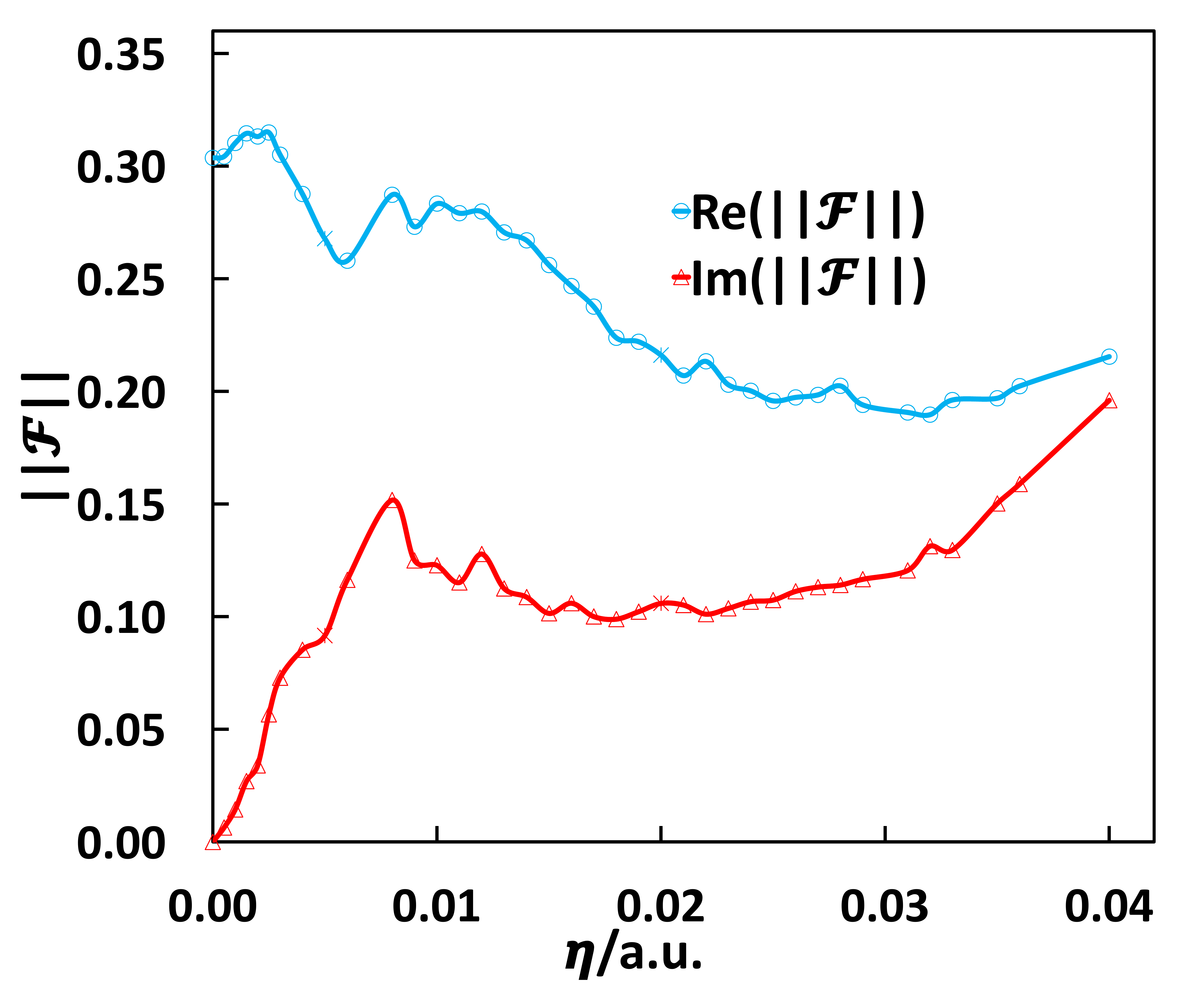}}
\captionsetup{justification=raggedright,singlelinecheck=false}
\caption{Non-adiabatic coupling force $h$ and derivative coupling 
$\mathcal{F}$ between the $^2$A$_g$ and $^2$A$_u$ resonances of 
fumaronitrile anion as a function of CAP strength $\eta$ computed 
at the equilibrium structure of the neutral molecule. 
See Sec. \ref{sec:compd} for explanation of the atom labels.}
\label{fig:rr}
\end{figure*}

As a final example, we studied the coupling between the $^2$A$_g$ 
and the $^2$A$_u$ resonance, which is mediated by vibrations of $a_u$ 
symmetry. This means that the only non-zero elements of $h$ and 
$\mathcal{F}$ are in $z$-direction. However, as illustrated by 
Fig. \ref{fig:rr}, there is no stabilization of $\text{Re}(h)$ or 
$\text{Im}(h)$ with respect to $\eta$ because the two resonances 
have different optimal CAP strengths. Fig. \ref{fig:rr} shows that 
the NAC force has significantly different character at the two 
respective $\eta_\text{opt}$ values of 0.005 a.u. and 0.02 a.u. 
At the lower value, the vector is dominated by the elements 
corresponding to movements of the carbon atoms, but their magnitude 
is much smaller at the higher value. We note that the lack of stability 
with respect to $\eta$ was also observed for transition dipole moments 
between two resonances computed with CAP-EOM-CCSD\cite{Jagau2016} 
and can be considered a fundamental weakness of CAP methods. Also 
noteworthy is that $\text{Re}(h)$ and $\text{Im}(h)$ are of similar 
magnitude, which is different from the coupling between a resonance 
and a bound state where the real part is dominant (see Sec. \ref{sec:res3}).

\section{Conclusions}\label{sec:conclusions}

We presented the theory and implementation of NAC vectors within 
the CAP-EOM-EA-CCSD framework, which is relevant for the study of 
non-adiabatic effects involving molecular temporary anions. Our 
approach is based on the Siegert representation, where electronic 
resonances are adiabatic states with complex energy and the resonance 
width is a local quantity. We also considered the 
connection of our approach to the treatment of NACs based on the 
Feshbach representation of electronic resonances.

The use of analytic gradient theory for CAP methods enables a treatment 
of polyatomic molecules that takes account of the full dimensionality 
of their PESs. We demonstrated this in a pilot application to anionic 
states of fumaronitrile, where we investigated the NACs between bound, 
resonance, and pseudocontinuum states. Our approach is most useful for 
evaluating couplings between a resonance and a bound state, where the 
results depend only weakly on the CAP strength. In contrast, couplings 
between two resonances depend more strongly on the CAP strength, which 
arises from a fundamental feature of CAP methods, namely, that optimal 
CAP parameters are specific to a particular resonance state. 

We see our work as a step towards the modeling of non-adiabatic 
effects involving metastable states in polyatomic molecules. Given the 
shortcomings of the CAP approach, it appears worthwhile to extend other 
approaches for electronic resonances to NACs. At the same time, we are 
convinced that our method is already useful in its present form as there 
are numerous other bound molecular anions besides fumaronitrile that have 
metastable excited states. NACs between them should leave fingerprints 
in spectroscopic experiments and we believe that NAC vectors computed 
with our method could help model them. 

%%%%%%%%%%%%%%%%%%%%%%%%%%%%
% Acknowledgements
%%%%%%%%%%%%%%%%%%%%%%%%%%%%
\section*{Acknowledgements}
This work was supported in Los Angeles by the U.S. National Science 
Foundation (CHE-2154482 to A.I.K.) and in Leuven by 
the European Research Council (ERC) under the European Union’s Horizon 
2020 research and innovation program (Grant No. 851766 to T.C.J.) and 
the KU Leuven internal funds (Grant C14/22/083).

\section*{Conflicts of interest}
The authors declare the following competing financial interest(s): 
A.I.K. is the president and a part-owner of Q-Chem, Inc.

\section*{Data availability}
The data that support the findings of this study are available within the article and the associated SI.   

%%%%%%%%%%%%%%%%%%%%%%%%%%%%
% Supplementary Material
%%%%%%%%%%%%%%%%%%%%%%%%%%%%
%\input{suppinfo}

%%%%%%%%%%%%%%%%%%%%%%%%%%%%
% References
%%%%%%%%%%%%%%%%%%%%%%%%%%%%
\bibliography{ref_kc}% Produces the bibliography via BibTeX.

\end{document}